\documentclass[a4paper,11pt]{article}
\usepackage{subcaption}     
\usepackage{jcappub}        
\usepackage{preamble}       
\usepackage{marginnote}

\fmtcountsetoptions{fmtord=level}

\newcommand{\maxstep}{\Delta \lambda_{\text{max}}}
\newcommand{\dLzmin}{0.002}
\newcommand{\dLzmax}{0.01}

\newcommand{\dLcoeff}[1]{d_L^{(#1)}}
\newcommand{\os}[1]{#1_o}
\newcommand{\dLcosmo}{d_{L \sssty{\rm (cosmo)}}}
\newcommand{\dLRT}{d_{L \sssty{\rm (RT)}}}
\newcommand{\DdL}{\Delta d_L}

\newcommand{\eH}{\mf{H}}
\newcommand{\eQ}{\mf{Q}}
\newcommand{\eR}{\mf{R}}
\newcommand{\eJ}{\mf{J}}
\newcommand{\eS}{\mf{S}}

\newcommand{\basis}{\xi}

\title{\boldmath Cosmography with $\Lambda$-Szekeres Models}

\author[a]{Morag Hills}
\affiliation[a]{School of Physical \& Chemical Sciences, University of Canterbury, Private Bag 4800, Christchurch 8140, New Zealand}
\author[b]{and Asta Heinesen}
\affiliation[b]{Department of Physics and Astronomy, Queen Mary University of London, UK}
\affiliation[b]{Niels Bohr Institute, Blegdamsvej 17, DK-2100 Copenhagen, Denmark}

\emailAdd{morag.hills@pg.canterbury.ac.nz}
\emailAdd{asta.heinesen@nbi.ku.dk}

\abstract{The cosmological tensions present in the $\Lambda$ cold dark matter model that have emerged and strengthened over recent years motivate model independent approaches to analysing data. Cosmography is useful for interpreting data in cosmology without imposing assumptions about the field equations of gravity or the matter content in the Universe. Some cosmography methods, denoted \emph{covariant cosmography}, go even further and stay  agnostic to the underlying space-time metric. {Due to their high level of generality,} covariant cosmography methods can incorporate {the} anisotropies and inhomogeneities in the observer's vicinity, and may in turn inform about the associated curvature of the relevant structures in our cosmic neighbourhood. Thus, covariant cosmography is a powerful model-independent tool for analysing cosmological data while also enabling the mapping of our local cosmic neighbourhood. In order to be able to explore the covariant cosmography framework to its fullest, it must be tested in tractable models and simulations. In this paper we derive the cosmography of luminosity distance to fourth order in redshift and investigate it in the special case of axially symmetric Szekeres models. {We compare the numerical results for the distance-redshift relations of synthetic observers placed within the Szekeres structures with the predictions from the cosmography, and comment on the found level of approximation of the cosmography in relation to other results in the litterature.} }

\begin{document}
\maketitle
\flushbottom

\section{Introduction}
\label{sec:intro}

    Cosmography is a route for achieving constraints {on} the expansion and curvature of the Universe without the specification of the field equations of gravity or the matter content of the space-time \citep{Weinberg:1972kfs,1985PhR...124..315E,Visser:2004bf}. 
    The methods of cosmography thus offer a truly model independent framework of analysing data that is complementary to parametrized methods that specify the dynamical model of the space-time in order to extract information about the space-time within that model scenario.
    Cosmography is useful for testing the consistency of the $\Lambda$ cold dark matter ($\Lambda$CDM) cosmology, and {is flexible enough} to {include virtually any} extension of it.

    With the still increasing list of possible theoretical scenarios for accounting for the cosmic tensions faced in the $\Lambda$CDM model, ranging from modified gravity scenarios to the addition of extra matter fields, cosmography presents an appealing framework for testing the extensions of the baseline $\Lambda$CDM model in a model-independent setup that can include a very diverse set of physical scenarios in the same setup.

    Cosmography has mainly been used for the purpose of constraining kinematic variables within the Friedmann-Lema\^itre-Robertson-Walker (FLRW) class of metrics, cf., e.g., \cite{Visser:2004bf}. However, cosmography is not limited to the FLRW geometries, and the methods of cosmography can in fact be extended in a, perhaps surprisingly, straightforward way to arbitrary space-time metrics{; see \cite{Kristian:1965sz,1985PhR...124..315E} for early work on such extensions.}
    Thus, cosmography can be used to extract model-independent information about the (anisotropic and inhomogeneous) degrees of freedom of the space-time geometry, with methods for this presented in \cite{Umeh:2013UCT,Clarkson:2011br,Heinesen:2020bej,Kalbouneh:2024szq}.
    A number of recent papers have demonstrated the feasibility of this aim within numerical relativity simulations \cite{Macpherson:2021gbh,Macpherson:2022eve,Adamek:2024hme}{, and have also taken the first steps to analyze real data \cite{Colin:2019opb,Dhawan:2022lze,Cowell:2022ehf,kalbouneh_anisotropic_2025}.  }

    Cosmography has been most widely investigated for measurements of luminosity distance and angular diameter distance, but can in principle be carried out for any observation that has a geometrical prediction, for instance position drift (once measured by \textit{Gaia}) \cite{Heinesen:2024npe} and redshift drift (measurable in the future by the Extremely Large Telescope and the Square Kilometre Array) \cite{Lobo:2020hcz,Rocha:2022gog,Heinesen:2021qnl}. In this work we focus on the cosmography for luminosity distance, but note that the drift effects are interesting complementary observables to consider for distinguishing different cosmological scenarios \cite{2010PhRvD..81d3522Q}.

    {One of the main challenges in applying cosmography to data is to control the level of approximation of the (truncated) cosmography and its convergence to the underlying distance-redshift relation.} 
    While the convergence properties of the FLRW cosmographic series for cosmic distances are reasonably well understood \cite{Cattoen:2007sk,capozziello_cosmographic_2017,capozziello_high-redshift_2020}, little work has been done in the literature for testing the convergence/level of approximation of space-times with anisotropies and/or inhomogeneities. 
    An exception is the recently published work \cite{Modan:2024txm,Sarma:2025yfw} in which analyses of the accuracy of the cosmographic approach within Lema\^itre-Tolman-Bondi (LTB) models were carried out.
    The LTB models are spherically-symmetric, and while they present good idealized models of isolated voids in the Universe, they are not ideal for describing large-scale over-densities of galaxies that are typically elongated in shape. 

    The Szekeres cosmological models are the most general exact solutions obtained for space-times sourced by dust. They are irrotational silent models universe models of Petrov type D, and they do not possess globally-defined Killing vectors.
    They thus present flexible models that may describe wall-like structures that are known to exist in the observed cosmic web of galaxies.
    Szekeres cosmological models are thus interesting tractable solutions of observational interest. The Hubble diagram has been investigated in a few studies employing ray tracing in Szekeres solutions, such as in \cite{2012MNRAS.419.1937M}. As far as we are aware, no paper has yet conveyed a detailed study of cosmography and convergence properties for Szekeres solutions, although 
    a recent paper 
    takes the first steps towards such investigations \cite{Celerier:2024dvs}.

    In this paper, we investigate the luminosity-distance cosmography for Szekeres space-times, with the purpose of systematically examining the approximation of the theoretical cosmography expressions relative to the exact luminosity distances computed by ray-tracing.
    In section \ref{sec:cosmo_form}, we review the general cosmographic expressions for luminosity distance derived previously, and we extend these results to include the fourth order term in the cosmographic series. In section \ref{sec:szek_models}, we review the Szekeres cosmological models for which we investigate cosmography in this paper.
    In section \ref{sec:rt_method}, we explain the ray tracing algorithm for calculating exact distances and redshifts in our codes.
    In section \ref{sec:results_main} we present and our results. We finally discuss our results in the context of recent work in the literature and conclude on our analysis in section \ref{sec:discussion}.

\vspace{3pt} 
\noindent
\underbar{Notation and conventions:}

We use units where $c=1$ and $G=1$, where $c$ is the speed of light and $G$ is the gravitational constant. 
Greek letters $\mu, \nu, \ldots$ are used for space-time
indices in a general coordinate system. 
The signature of the space-time metric $g_{\mu \nu}$ is $(- + + +)$ and the covariant derivative $\nabla_\mu$ is the usual metric-preserving Levi-Civita connection. 
Vectors, $V^\mu$, are sometimes written in the basis free notation $\boldsymbol{V}$ for the sake of compactness. 
Covariant derivatives of vectors are ocassionaly written as $\cdd{V^\mu}{\nu} \equiv \nabla_\nu V^\mu$.

\section{Covariant Cosmography Formalism}\label{sec:cosmo_form}

\subsection{Fundamental congruences}\label{sec:fundamental_congruences}

We shall now formulate a \emph{covariant luminosity distance cosmography}, \ie, a Taylor series expansion of luminosity distance in redshift, valid in any Lorentzian space-time where light is travelling on non-caustic null geodesics and where the observers and emitters (of light and gravitational waves) of interest are collectively described by a time-like congruence of worldlines.
The derivations follow those of \cite{Heinesen:2020bej}, and the reader is referred to this paper for details.

There are two congruences of interest to be considered in obtaining the generalised luminosity distance Hubble law. The first of these congruences is the time-like congruence, belonging to the reference observers and emitters, which is described by the tangent vector field $\bs{u}$ - the 4-velocity of the observers and emitters. The second of these congruences is the light-cone forming irrotational\footnote{The irrotational property follows directly from the requirement that the congruence represents the lightcone of an observer.} geodesic congruence of photons, described by the affinely parametrised tangent null vector $\bs{k}$. This vector corresponds to the 4-momentum of the photons emitted and observed between the members of the time-like observer-emitter congruence.

The photon 4-momentum $\bs{k}$ can be decomposed into parallel and orthogonal parts to the 4-velocity $\bs{u}$ as follows:
\begin{equation}\label{eq:kmuE}
    k^\mu = E\qty(u^\mu - e^\mu)\,, \quad E \equiv -k^\mu u_\mu\,,
\end{equation}
where the unit vector field $\bs{e}$ is orthogonal to $\bs{u}$, and indicates the spatial propagation direction of a photon measured by the observers comoving with $\bs{u}$. 
As such, one can consider $\bs{e}$ as a radial unit vector when evaluated at the vertex of the observer's lightcone. The function $E$ is the energy of the photon congruence inferred by the observers comoving with $\bs{u}$, and can be fixed through the freedom in rescaling of the affine parametrisation of $\bs{k}$ \cite{ellis_relativistic_2012}.

The evolution of the energy function $E$ along the null geodesics can be formulated in terms of projected dynamical variables of $\bs{u}$ in the following way: 
\begin{equation}\label{eq:dEdl}
    \dv{E}{\lambda} = -k^\mu\nabla_\mu\qty(k^\nu u_\nu) = -E^2\,\mf{H}\, , 
\end{equation}
{where we have defined the derivative along the photon 4-momentum $\dv{}{\lambda} \equiv k^\mu \partial_\mu$. }
By taking the covariant derivative of $E$ with respect to the affine parameter $\lambda$ along the geodesics, we obtain the parameter 
\begin{equation}\label{eq:effh}
    \mf{H} \equiv \frac{1}{3}\Theta - e^\mu a_\mu + e^\mu e^\nu \sigma_{\mu\nu}\,,
\end{equation}
which is a truncated multipole series in the spatial unit vector $\vb*{e}$, wherein $\Theta$ is the volume expansion rate of the congruence, $\sigma_{\mu \nu}$ is the volume shear rate describing its symmetric anisotropic deformation, and $a^\mu = u^\mu\cd{\nu}u^{\mu}$ is the 4-acceleration of the field. The coefficients of \cref{eq:effh} are expressed in terms of the kinematic decomposition of the matter frame as
\begin{align} \label{def:expu}
        & \nabla_{\nu}u_{\mu}  = \frac{1}{3}\Theta\, h_{\mu \nu }+\sigma_{\mu \nu} + \omega_{\mu \nu} - a_\mu u_\nu  \, , \nonumber \\ 
        & \Theta \equiv \nabla_{\mu}u^{\mu} \, , \quad \sigma_{\mu \nu} \equiv h^\beta_{\, ( \mu} h^\alpha_{\, \nu ) } \nabla_{\alpha} u_{\beta} - \frac{1}{3} \Theta\, h_{\mu \nu} \, , \quad  \omega_{\mu \nu} \equiv h^\beta_{\, [ \mu} h^\alpha_{\, \nu ] } \nabla_{\alpha} u_{\beta}\,, 
\end{align} 
where $\omega_{\mu \nu}$ describes the rotational deformation of the congruence and $h_{\mu\nu}=g_{\mu\nu} + u_\mu u_\nu$ is the projection tensor to the matter frame, with round parentheses implying symmetrisation over the selected indices and square brackets implying anti-symmetrisation.

The function $\mf{H}$, as given by \cref{eq:effh}, which we denote the \emph{effective Hubble parameter} describes the rate of change of photon energies with affine distance in general space-times, and thus generalise the isotropic Hubble parameter $H$ in FLRW spacetime. In the following \cref{sec:distmeasures}, we {shall see that $\mf{H}$ indeed defines the leading order distance-redshift proportionality law in the general spacetime setting, thus generalizing the wellknown Hubble law. We will also} demonstrate how we can further obtain the equivalent \emph{effective cosmographic parameters} for the FLRW deceleration $q$, jerk $j$ and snap $s$ parameters, by taking successively higher order derivatives of \cref{eq:effh} and \cref{def:expu}.  

\subsection{Distance-redshift measures}\label{sec:distmeasures}

\newcommand{\dLflrwcoeffA}{\dLcoeff{1} = \frac{1}{\os{H}}}
\newcommand{\dLflrwcoeffB}{\dLcoeff{2} = \frac{1 - \os{q}}{2\os{H}}}
\newcommand{\dLflrwcoeffC}{\dLcoeff{3} = \frac{-1 + 3\os{q}^2 + \os{q} -\os{j} + \Omega_k}{6\os{H}}}
\newcommand{\dLflrwcoeffD}{\dLcoeff{4} = \frac{2 - 2\os{q} - 15\os{q}^2\qty(1 + \os{q}) + 5\os{j} + 10\os{q}\os{j} + \os{s} - 2\Omega_k\qty(1 + 3\os{q})}{24\os{H}}}

The angular diameter distance $d_A$ of an observed source, which measures the physical extent of an object relative to its angular extent subtended on the observer's sky, is given by
\begin{equation}
    d_A = \sqrt{\frac{\delta A}{\delta\Omega}}\,,
\end{equation}
where $\delta A$ is the cross-sectional area of the source (perpendicular to the 4-velocity $\bs{u}$ and 4-momentum $\bs{k}$ of the source congruence) and $\delta\Omega$ is the observed angular extent. As the light bundle propagates through space, the angular diameter distance will evolve according to the expansion rate of the null congruence, $\hat{\theta} = \cd{\mu}{k^\mu}$, as follows
\begin{equation}\label{eq:ddA}
    \dv{d_A}{\lambda} = \hlf\hat{\theta}d_A\,.
\end{equation}
{It} follows from the Sachs equations, defining the deformation rates of the expansion $\hat{\theta}$ and shear $\hat{\sigma}_{\mu\nu}$ of the null congruence, that 
\begin{subequations}
    \begin{gather}
        \dv{\hat{\theta}}{\lambda} = -\hlf\hat{\theta}^2 - \hat{\sigma}_{\mu\nu}\hat{\sigma}^{\mu\nu} - k^\mu k^\nu R_{\mu\nu}\,, \label{eq:sachstheta}\\
        {\frac{\rm{D} \hat{\sigma}_{\mu\nu}}{\rm{d}\lambda} }  = -\hat{\theta}\hat{\sigma}_{\mu\nu} - {p^\gamma\!}_\mu\,{p^\sigma\!}_\nu\,k^\alpha k^\beta C_{\gamma\alpha\sigma\beta}\,, \label{eq:sachssigma}
    \end{gather}
\end{subequations}
{where we have defined the covariant derivative along the photon 4-momentum $\frac{\rm{D}}{\rm{d}\lambda} \equiv k^\mu \nabla_\mu$, and where  $C_{\gamma\alpha\sigma\beta}$ is the Weyl curvature tensor of the space-time. } 
{We} can express the second derivative of the angular diameter distance as
\begin{equation}\label{eq:d2dA}
    \dv[2]{d_A}{\lambda} = -\qty(\hat{\sigma}^2+\frac{1}{2}k^\mu k^\nu R_{\mu\nu})\,d_A\,
\end{equation} 
by substituting (\ref{eq:sachstheta}) into the derivative of (\ref{eq:ddA}). 
In the above, $p_{\mu\nu}=g_{\mu\nu} + u_\mu u_\nu - e_\mu e_\nu$ is the projection tensor onto the 2-dimensional \emph{screen spaces} that are orthogonal to $u^\mu$ and $e^\mu$. The first term of (\ref{eq:d2dA}), containing the scalar shear, $\hat{\sigma}^2 \equiv \frac{1}{2} \hat{\sigma}^{\mu\nu} \hat{\sigma}_{\mu\nu}$, of the null congruence is responsible for Weyl focusing of the congruence, whereas the second term is responsible for Ricci focusing of the congruence. 

The luminosity distance, $d_L$, to a source is defined in terms of the source's bolometric luminosity $L$ at the source and measured bolometric flux $F$ by the observer: 
\begin{equation}
    d_L = \sqrt{\frac{L}{4\pi F}}\,.
\end{equation}
One can relate the angular diameter distance to the luminosity distance $d_L$ separating the source and observer through Etherington's reciprocity theorem, assuming that photon conservation holds, as 
\begin{equation}\label{eq:dLzdA}
    d_L = (1+z)^2d_A\, , 
\end{equation}
where the redshift is defined as 
\begin{equation}\label{eq:red}
 z \equiv \frac{E}{E_o} - 1 \, , 
\end{equation}
where $E_o$ is the energy of the photon as measured by the observer at the vertex of the past lightcone. 

Using the known expressions for the derivatives of $E$ and $d_A$ with respect to $\lambda$ derived in the above equations \eqref{eq:dEdl}, \eqref{eq:ddA}, and \eqref{eq:d2dA}, and together with the definition of redshift in \eqref{eq:red}, we can write the Taylor series expansion of $d_L$ in $z$ 
\citep[see][for details]{Heinesen:2020bej} 
\begin{equation}\label{eq:series}
    d_L(z) =  d_L^{(1)} z   + d_L^{(2)} z^2 +  d_L^{(3)} z^3 + \mathcal{O}( z^4),
\end{equation} 
wherein the coefficients may be expressed in terms the effective cosmographic parameters evaluated at the vertex point of the congruence, \ie, at the point of observation as denoted by $o$ subscripts, as follows:
\begin{subequations}\label{eqs:dLcoeffs}
    \begin{gather}
        {d_L}^{(1)} = \frac{1}{\os{\eH}}\,, \\
        {d_L}^{(2)} = \frac{1-\os{\eQ}}{2\os{\eH}}\,, \\
        {d_L}^{(3)} = \frac{-1+3\os{\eQ}^2+\os{\eQ}-\os{\eJ}+\os{\eR}}{6\os{\eH}}\,, \\
        {d_L}^{(4)} = \frac{2-2\os{\eQ}-15\os{\eQ}^2-15\os{\eQ}^3+5\mf{J}_+10\os{\eQ}\os{\eJ}+\os{\eS}-2\os{\eR}\qty(1+3\os{\eQ})}{24\os{\eH}} \, 
    \end{gather}
\end{subequations}
with 
\begin{subequations}\label{eqs:effparams}
    \begin{gather}
        \eQ \equiv -1-\frac{1}{E}\frac{\dv{\eH}{\lambda}}{\eH^2}\,, \label{eq:effQ}\\
        \eR \equiv 1 + \eQ - \frac{1}{2E^2}\frac{k^\mu k^\nu R_{\mu\nu}}{\eH^2}\,, \label{eq:effR}\\
        \eJ \equiv \frac{1}{E^2}\frac{\dv[2]{\eH}{\lambda}}{\eH^3} - 4\eQ - 3\,, \label{eq:effJ}\\
        \eS = \frac{1}{E^3}\frac{\dv[3]{\eH}{\lambda}}{\eH^4} + \frac{1}{E^3}\frac{\dv{\lambda}\qty(k^\mu k^\nu R_{\mu\nu})}{\eH^3} - 8\eR + 4\eQ^2 + 35\eQ + 9\eJ + 23\, .  \label{eq:effS}
    \end{gather}
\end{subequations}
{Immediately, we find the coefficients in \eqref{eqs:dLcoeffs}} to be analogous to the typical presentation of the FLRW series expansion coefficients: $\dLflrwcoeffA$, $\dLflrwcoeffB$, $\dLflrwcoeffC$, and $\dLflrwcoeffD$, where $\Omega_k$ 
is the FLRW spatial curvature density. 
{We thus identify the parameters in} \cref{eqs:effparams} as \emph{effective cosmographic parameters}, in addition to the previously identified effective Hubble parameter $\eH$ defined by (\ref{eq:effh}): the effective `deceleration' \cref{eq:effQ}, `curvature' \cref{eq:effR}, `jerk' \cref{eq:effJ} and `snap' \cref{eq:effS} parameters, respectively. 
The \ord{3} order expansion agrees with that already derived in \cite{Heinesen:2020bej}, and our results extend the work of \cite{Heinesen:2020bej} with the added \ord{4} order term. 
Kalbouneh \etal \cite{kalbouneh_expanding_2025} have previously calculated the \ord{4} order cosmographic series using a slightly different version of the formalism presented in this paper. We have compared our results to theirs, with agreement up to third order.

Computing the effective cosmographic parameters in \eqref{eqs:effparams} requires taking appropriate covariant derivatives (with respect to $\lambda$) of $\mf{H}$ \cref{eq:effh}. 
To investigate the \ord{4} order cosmography, we thus computed all covariant derivatives up to \ord{3} order of the variables $\Theta$, $\sigma_{\mu\nu}$ and $e^\mu$ entering in $\mf{H}$ (\ref{eq:effh}). Additionally, for \cref{eq:effS}, we computed the \ord{1} order covariant derivative of $R_{\mu\nu}$. 
For convenience, we provide expressions for the derivatives of $e^\mu$ up to \ord{3} in appendix~\ref{sec:derivdirection}.

\subsection{Alternative approximants}\label{sec:pade}

The cosmographic Taylor series expansion in powers of redshift typically converge for $z \leq 1$ \cite{cattoen_hubble_2007}, at least when one considers the standard isotropic cosmographic expansion of the Friedmann solutions. 
Various solutions have been proposed to extend the convergence radius of the standard cosmographic series expansion for $d_L$ and $d_A$ above $z=1$, including parameterisations by auxiliary variables, Chebyshev polynomials and Pad\'e approximants (\eg \cite{cattoen_hubble_2007,busti_is_2015,capozziello_high-redshift_2020,nesseris_comparative_2013,capozziello_cosmographic_2017,sapone_curvature_2014,gruber_cosmographic_2014}) -- each with their own benefits and shortcomings. Pad\'e approximants can also help increase the convergence radii for cosmographies performed within realistic anisotropic space-times \cite{adamek_towards_2024}. However, these approximants are yet to be fully explored in the context of inhomogeneous models.  
In this paper, we consider a range of appropriate Pad\'e approximants \cite{baker_pade_1996,press_numerical_1992} as alternatives to the standard Taylor series expansion for $d_L$ \eqref{eq:series}. Given a Taylor series expansion of the form
\begin{equation}\label{eq:padebasic}
    f(x) = \sum\limits_{k=0}^K c_k x^k\,,
\end{equation}
for a function $f(x)$, the corresponding Pad\'e approximant $P(x)$ of order $[M/N]$ is the rational polynomial
\begin{equation}\label{eq:padeformula} 
   P(x)_{M/N} = \frac{\sum\limits_{i=0}^M a_i x^i}{1 + \sum\limits_{j=1}^{N}b_j x^j}\,,
\end{equation}
where $M$ and $N$ give the orders of the numerator and denominator, respectively, and $M+N\le K$. Furthermore, through setting each order of \cref{eq:padebasic} and \cref{eq:padeformula} equal to each other, one can solve for the coefficients $a_i$ and $b_j$ within \cref{eq:padeformula} using the known coefficients $c_k$ of the Taylor series. 

\section{Szekeres cosmological models}\label{sec:szek_models}

\subsection{Line element and field equations}

The class of Szekeres models considered in this paper contain only a pressureless, irrotational dust source with rest-mass density $\rho$ and cosmological constant $\Lambda$, with the metric defined in synchronous comoving coordinates. The Szekeres line element in its `projective coordinate' $(p,q)$ representation is
\begin{equation}\label{eqn:szek_pq_metric}
    \dd{s}^2 = -\dd{t}^2 + \frac{\qty(R'-R\,\mc{E}'/\mc{E})^2}{\epsilon-k}\,\dd{r}^2 + \frac{R^2}{\mc{E}^2}\qty(\dd{p}^2+\dd{q}^2)\,,
\end{equation}
where primes denote a partial derivative with respect to the radial coordinate, $r$. Note that $r$ does not correspond to the radial distance from the origin of the model, as it otherwise does in the LTB models, because the non-concentric arrangement of Szekeres shells means that each shell's origin need not coincide. The parameter $\epsilon=+1,0,-1$ delineates between the quasi-spherical, quasi-planar and quasi-pseudo-spherical subtypes; we consider the well-studied $\epsilon=+1$ case here. The function $R=R(t,r)$ is the proper area radius of a given shell, and $k=k(r)$ is the local spatial curvature parameter.

The function $\mc{E}=\mc{E}(r,p,q)$ in Szekeres metric \cref{eqn:szek_pq_metric} describes the deviation of a given Szekeres model from its corresponding LTB model limit, and reads 
\begin{equation}\label{eqn:e_func_pq}
    \mc{E}(r,p,q) = \frac{\qty(p-P(r))^2 + \qty(q-Q(r))^2 + \epsilon S(r)}{2S(r)}\,,
\end{equation}
where the dipole functions $S(r)$, $P(r)$ and $Q(r)$ characterise the dipole asymmetry. In appendix~\ref{sec:metricderived}, we provide the non-zero Chrisfoffel symbols and Ricci tensor components of the Szekeres models in projective coordinates, as well as the derivatives of $\mc{E}$ which enter the Ricci tensor (see also \cite{Celerier:2024dvs} for useful curvature identities). 

The Einstein equations of the Szekeres model are 
\begin{equation}
\label{eq:EE}
    R_{\mu\nu} = 8\pi\qty(T_{\mu\nu}-\hlf Tg_{\mu\nu})-\Lambda g_{\mu\nu} = 8\pi\rho\qty(u_\mu u_\nu+\hlf g_{\mu\nu}) - \Lambda g_{\mu\nu}\,, 
\end{equation}
where $\rho$ is the rest-mass density of the source dust. They reduce to two independent equations, the first being the time evolution of $R(t,r)$,
\begin{equation}\label{eq:evo}
    \dot{R}(t,r)^2 = \frac{2M(r)}{R(t,r)} - k(r) + \frac{1}{3}\Lambda R(t,r)^2\,,
\end{equation}
where the function $M(r)$ is the active gravitational mass contained within a constant-$r$ shell in the quasi-spherical case, which is responsible for generating the gravitational field\footnote{This is distinct from the integrated mass of all dust particles comprising the matter content of the model \cite{bondi_spherically_1947}.}. Integrating (\ref{eq:evo}), we further obtain
\begin{equation}\label{eq:bangt}
    t-t_B(r) = \int^{R(t,r)}_0\qty(\frac{2M(r)}{\Tilde{R}} - k(r) + \frac{1}{3}\Lambda\Tilde{R}^2)^{-1/2}\dd{\Tilde{R}}\,,
\end{equation}
where $t_B(r)$ is the `bang-time' at a given $r$\footnote{The radially dependent bang-time function $t_B(r)$ describes the time at which a given constant-r Szekeres shell emerges from the $R(t_B,r) = 0$ singularity, as each shell evolves independently of each other.}.  The second equation we obtain from the Einstein equations is the mass density equation
\begin{equation}\label{eq:szekrho}
    4\pi\rho(t,r,p,q) = \frac{M'-3M\mc{E}'/\mc{E}}{R^2(R'-R\mc{E}'/\mc{E})}\, , 
\end{equation}
where the coordinate dependence of the right hand side of the equation is implicit for the sake of compactness.
It is evident here that the factors of $\mc{E}'/\mc{E}$ control the dipole distribution of matter.
The Szekeres line element (\ref{eqn:szek_pq_metric}) can be recast in spherical coordinates via a Riemannian stereographic projection
\begin{align}\label{eq:szek_sph_metric}
    \dd{s}^2 = & -\dd{t}^2 + \frac{1}{1-k}\qty[R'-\frac{R}{S}\qty(S'\cos\theta+\mc{N}\sin\theta)]^2\dd{r}^2 \nonumber                                                     \\
               & +\qty[\frac{S'\sin\theta+\mc{N}(1-\cos\theta)}{S}]^2R^2\dd{r}^2 + \qty[\frac{\qty(\partial_\phi\mc{N})(1-\cos\theta)}{S}]^2R^2\dd{r}^2 \nonumber         \\
               & -2\frac{S'\cos\theta+\mc{N}\sin\theta}{S}R^2\dd{r}\dd{\theta} + 2\frac{\qty(\partial_\phi\mc{N})(1-\cos\theta)}{S}R^2\sin\theta\dd{r}\dd{\phi} \nonumber \\
               & + R^2\qty(\dd{\theta}^2 + \sin^2\theta\dd{\phi}^2)
\end{align}
where $\mc{N}(r,\phi) = \qty(P'\cos\phi + Q'\sin\phi)$. As the Szekeres models we examine in this paper are axially symmetric, we restrict $S'(r)$ to be the only non-zero dipole function, so that $\mc{N}(r,\phi) = 0$. The models are thus symmetric about the axis defined by $S(r)$ with the dipole position only varying in the $\theta$ coordinate, as there will be no $\phi$ dependence in $\mc{E}'/\mc{E}$ when $P'=Q'=0$. 

The projective coordinate representation of the metric (\ref{eqn:szek_pq_metric}) is straight-forward to use in deriving various products of the metric, owing largely to its diagonalised form. 
However, for the purpose of our specific numerical ray tracing scheme {that we employ in this paper}, the spherical coordinates of \eqref{eq:szek_sph_metric} are convenient and avoid coordinate singularities along the symmetry axis \cite{buckley_physical_2020}. 
Thus, in the present work, we use the spherical coordinate form of the Szekeres metric for ray tracing, while we use the simpler projective coordinates for the computation of the covariant cosmographic parameters in \eqref{eqs:effparams}.

\subsection{Model parameterisation}

We define the background FLRW model that our Szekeres models approach at large radii by setting the cosmological parameters to the \textit{Planck} parameter estimates \cite{Planck:2018vyg},
\begin{equation}\label{eq:flrw_params}
    \qty(\Omega_{m0},\,\Omega_{\Lambda0},\,\Omega_{k0},\,h) = (0.315,\,0.685,\,0.0,\,0.673)\,,
\end{equation}
where $\Omega_{m0}$, $\Omega_{\Lambda0}$ and $\Omega_{k0}$ are the present-time Friedmann values of the matter, cosmological constant and spatial curvature densities, respectively. Within our simulations, the big-bang singularity is made simultaneous for all comoving observers through fixing $t_B(r)=0$ for all $r$ in \cref{eq:bangt}. We determine the present age of the universe via the following analytic solution of the background Friedmann equation
\begin{equation}\label{eq:t0}
    t_0 = \frac{1}{3H_0\sqrt{\Omega_{\Lambda0}}}\ln\qty(\frac{1+\sqrt{\Omega_{\Lambda0}}}{1-\sqrt{\Omega_{\Lambda0}}})\,,
\end{equation}
as we have $\Omega_{k0}=0$ for our flat FLRW background model. The function $R(t,r)$ is set to coincide with the coordinate-$r$ at the present time $t_0$, such that $R(t_0,r)=r$, which is used as the initial condition for \cref{eq:evo}. 

In the axially symmetric Szekeres models that we investigate, the dipole functions are defined as,
\begin{equation}\label{eq:spqfuncs}
    P(r)=0, \qquad Q(r)=0, \qquad S(r) = \qty(\frac{r}{1\,\mpc})^\alpha\,\mpc\,,
\end{equation}
where the parameter $\alpha$ is introduced to effectively control the strength of the dipole in the density distribution. 

\begingroup
\renewcommand{\arraystretch}{1.2}
\begin{figure}[h!]
    \centering
    \begin{floatrow}
        \tablebox{\begin{tabular}{ccccc}
            \toprule
                    & $\alpha$ & $\delta_0$ & $\Delta r$ \\
            \midrule
            Model 1 & $0.5$    & $-0.8$     & $40\,\hmpc$ \\
            Model 2 & $0.5$    & $-0.8$     & $80\,\hmpc$ \\
            \bottomrule
        \end{tabular}}{\caption{Parameter values for model 1 and model 2. These parameters are used within \cref{eq:expdelta} to control the behaviour of the Szekeres mass profile \cref{eq:massfunc}.}
        \label{tab:model_params}}
        \hfill
        \tablebox{\begin{tabular}{ccccc}
                \toprule
                           & $\ffrac{r_{\rm obs}}{\Delta r}$ & $\theta_{\rm obs}$ & $\phi_{\rm obs}$ \\
                \midrule
                Observer 1 & $0.875$    & $0.7\,\pi$         & $0.5\,\pi$       \\
                Observer 2 & $0.875$    & $0.3\,\pi$         & $0.5\,\pi$       \\
                Observer 3 & $0.625$   & $0.7\,\pi$         & $0.5\,\pi$       \\
                Observer 4 & $0.625$   & $0.3\,\pi$         & $0.5\,\pi$       \\
                \bottomrule
            \end{tabular}}{\caption{Selected observer positions in Szekeres spherical coordinates.}\label{tab:obs_params}}
    \end{floatrow}
\end{figure}
\endgroup

\begin{figure}[ht]
    \centering
    \includegraphics[width=0.8\textwidth]{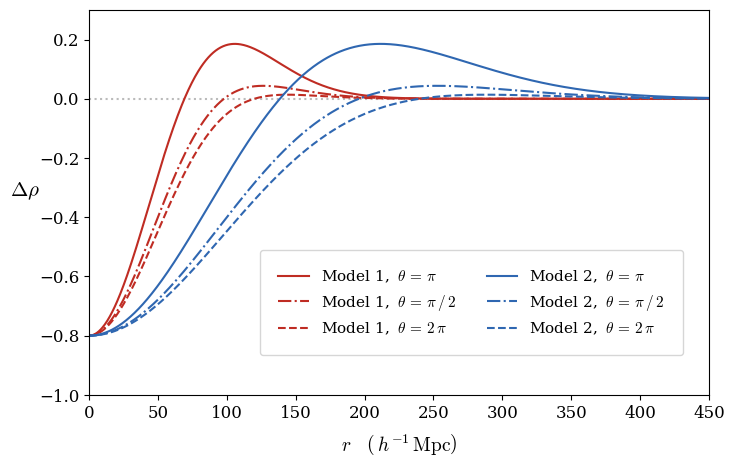}
    \caption{The initial present time 2D density profiles of model 1 and model 2, normalised to the background FLRW density $\Delta\rho = (\rho-\bar{\rho})/\bar{\rho}\,$. These profiles are plotted from the coordinate centre of the Szekeres model, with $r$ being the radial distance from the coordinate centre. In terms of the angular Szekeres coordinate $\theta$, solid $\theta=\pi$ lines corresponding to the angular position of the maximum over-density magnitude, while dashed $\theta=2\pi$ lines correspond to the angular position of the minimum over-density magnitude. Dash-dotted lines give the over-density magnitude located $90^\circ$ from these extrema. The parameter values of each profile are given in table \ref{tab:model_params}.}
    \label{fig:2d_density_profiles}
\end{figure}

We parametrise the mass distribution of the Szekeres solution as follows  
\begin{equation}\label{eq:massfunc}
    M(r) = M_0(r)+\delta M(r) 
\end{equation}
where the homogeneous mass profile $M_0(r)$ of the FLRW background and the deviation of the mass profile from the background $\delta M(r)$ are respectively given by
\begin{equation}\label{eq:massflrw}
    M_0(r) = \frac{4}{3}\pi\bar{\rho}\qty(t_0)r^3 = \frac{1}{2}\Omega_{m0}{H_0}^2r^3\,, \qquad \delta M(r) = \frac{1}{2}\Omega_{m0}{H_0}^2r^3 \delta(r) \, . 
\end{equation}
We finally construct our Szekeres models by specifying their density profiles (\ref{eq:szekrho}) through the function $\delta(r)$ entering in equation (\ref{eq:massflrw}). We select the following form:
\begin{equation}\label{eq:expdelta}
    \delta(r) = \delta_0\,\exp\qty[-\qty(\frac{r}{2\,\Delta r})^2]\,,
\end{equation}
wherein the parameters $\delta_0$ and $\Delta r$ allow one to tune the characteristics of the resulting density profile. Setting the local perturbation parameter $\delta_0 \in [-1,0 )$ produces an under-density or void centred at the origin $r=0$. 
The parameter $\Delta r$ is used to control the steepness of the void density profile, \ie, the transition rate between the under-dense and over-dense regions of the model. 
The profile \cref{eq:expdelta} thus allows us to describe a local under density (with a corresponding over density) at varying characteristic radii. 
We summarise the models and observer positions that we investigate in our analysis in \cref{tab:model_params,tab:obs_params}. 
The density profiles for the models in \cref{tab:model_params} are shown in \cref{fig:2d_density_profiles} for selected directions away from the centre of the under density. 

\section{Ray tracing methodology}\label{sec:rt_method}

\subsection{Null geodesic equations}

Performing a thorough analysis of the covariant cosmographic expansions requires one to compute the exact quantities of interest, \eg the redshift and angular diameter distance, as seen by a given fiducial observer within our Szekeres models for comparative means. Ray tracing analyses in both Szekeres and LTB models have been performed for various purposes in past research (see, \eg \cite{bolejko_differential_2016,koksbang_studying_2015,Macpherson:2022eve}). In this paper, we adopt a similar ray tracing scheme to that used in \cite{bolejko_differential_2016}, and later developed in \cite{dam_inhomogeneous_2016,hills_ray_2022}. This scheme is centred around the numerical integration of the null geodesic equations,
\begin{equation}\label{eq:nullgeo}
  k^\mu \nabla_\mu k^\gamma =  \dv{k^\gamma}{\lambda} + \chris{\gamma}{\mu}{\nu}\,k^\mu k^\nu = 0\,,
\end{equation}
where $\chris{\gamma}{\mu}{\nu}$ are the Christoffel symbols and $\lambda$ is the affine parameter along the light ray and the tangent null vectors $k^\mu$ are defined as 
\begin{equation}
\label{eq:kdef}
     k^\mu = \dv{x^\mu}{\lambda} \,.
\end{equation}
Due to the complexity of the spherical coordinate representation of the Szekeres metric, we do not include their precise forms in this paper. However, the non-zero Christoffel symbols in projective coordinates are presented for reference in appendix \ref{app:projchris} for completeness. The primary aim of our ray tracing procedure is to obtain a given observable quantity, such as the luminosity distance $d_L$ or angular diameter distance $d_A$, as observed at a fixed target redshift $z$ or vice versa, by backwards propagating the null geodesic equations through spacetime until the target is achieved.

For the numerical ray tracing simulations undertaken in this investigation, we use the \textsc{healp}ix (Hierarchical Equal Area isoLatitude Pixelisation) scheme via the \textsc{Python} and \textsc{Julia} implementations \texttt{healpy} \cite{zonca_healpy_2019} and \texttt{healpix.jl} \cite{2021ascl.soft09028T}, respectively. We numerically solve the system of ODEs for the null geodesic equations \cref{eq:nullgeo} using the \texttt{tsit5()} method provided by the \texttt{DifferentialEquations.jl} package \cite{rackauckas_differentialequationsjl_2017}, which is an implementation of Tsitouras' 5th order explicit Runge-Kutta method \cite{tsitouras_rungekutta_2011}.

For each pixel on the observer's \textsc{HEALPix} sky, we \emph{backwards propagate} a light ray until the redshift $z$ is achieved, and then evaluate the luminosity distance at the corresponding event. We note that in our description of the ray tracing method that follows, we define our null vectors $k^{\hat{\alpha}}_{\rm obs}$ at the observer to be future-pointing. However, our codes are constructed using past-pointing null vectors, which is simply a choice of convention. For further technical details regarding the stability and convergence of the numerical implementation; see appendix~\ref{app:techdetails}.  

\subsection{Ray initialisation}

In order to determine the redshift $z$ of the source, we propagate rays in from the observer to the source by effectively tracking the solutions of eight ODEs, corresponding to \eqref{eq:nullgeo} and \eqref{eq:kdef} in the variables 
$k^\mu = \qty(k^t, k^r, k^\theta, k^\phi)$ and the ray position vector $x^\mu = \qty(x^t,x^r,x^\theta,x^\phi)$ at each step of the numerical integration routine. 
At the observer, we initialise $k^t=1$ via our choice of affine parameter so that the redshift is given by 
\begin{equation}\label{eq:ktz}
    1 + z = \frac{\qty(k^\mu u_\mu)_{\rm src}}{\qty(k^\mu u_\mu)_{\rm obs}} = \frac{k^t_{\rm src}} {k^t_{\rm obs}} = k^t_{\rm src}\,,
\end{equation}
where we note that for the comoving (geodesic) Szekeres observer we simply have $u_\mu = (-1,0,0,0)$. 
From the decomposition \cref{eq:kmuE} of the null vector $k^\mu$, the spatial unit vector $\bs{e}^{\hat{\alpha}} = (0, \vu*{n})$ is equivalent to the direction on the observer's sky in which a source is observed. The direction vector can be converted in terms of equatorial coordinates $\qty(\ell,b)$ on the sky, (or as in our numerical implementation, as individual indexed pixels on a \textsc{heal}pix map). By then numerically determining $k^t_{\rm src}$ through (backwards) propagation of the ray from observer to source, via integration of the null geodesic equations \cref{eq:nullgeo} that one may obtain the source's $z$ in a given Szekeres model. 

The angular diameter distance is computed via the focusing equation (\ref{eq:d2dA}). We set the Weyl focusing term $\propto \hat{\sigma}^2$ to zero in this analysis\footnote{For full numerical solutions including the Weyl term, see, e.g. \cite{Modan:2024txm, Macpherson:2022eve}}, under the assumption that it is subdominant. (See, \eg, \cite{bolejko_differential_2016,bolejko_ricci_2012} for a justification of this.)
Under this approximation, the focusing equation \cref{eq:d2dA} becomes
\begin{equation}
    \dv[2]{d_A}{\lambda} + 4\pi\rho\,\qty(k^t)^2d_A = 0\, ,
\end{equation}
which follows from the Einstein equations \eqref{eq:EE}. 
Lastly, it remains in the initialisation procedure to set up the initial null vector of each ray in the local orthonormal frame. For each angular direction of observation $\qty(\ell,b)$ at the observer, where in this case $b$ is the co-latitude and $\ell$ the longitude measured in radians, we first determine the Cartesian direction vector $\vu*{n}$ in the local inertial frame 
\begin{equation}\label{eq:localframe}
    \hat{n}^x = \cos{b}\cos{\ell}\,,\quad \hat{n}^y = \cos{b}\sin{\ell}\,,\quad \hat{n}^z = \sin{b}\,,
\end{equation}
such that the initial null vector is then given by $k^{\hat{\alpha}}_{\rm obs} = \qty(1,-\vu*{n})$
in this choice of coordinates. We initialise our ray null vectors in the Szekeres spherical coordinates of \cref{eq:szek_sph_metric} by contracting the local orthonormal basis tetrad ${e^\mu}_{\!\hat{\alpha}}$ with $k^{\hat{\alpha}}$ at the observer position, so that in these coordinates: $k^\mu_{\rm obs} = \left.{e^\mu}_{\!\hat{\alpha}} k^{\hat{\alpha}}\right|_{\rm obs}$. The details of this normalisation procedure are described in further detail in appendix~\ref{app:nullvecinit}.

\section{Results}\label{sec:results_main}
\newcommand{\figcapz}{This is presented as full sky maps for hypothetical sources at five equally spaced {redshifts} between $z=\dLzmin$ and $z=\dLzmax$, with the observer located at $z=0$.}
\newcommand{\figcapdLdiffobsone}[1]{Variation of $\Delta d_L$ (\ref{eq:dLdiff}) up to \ord{4} order in the covariant cosmographic expansion estimate of $d_L$ versus the exact $d_L$ obtained via ray tracing for observer 1 in model {#1}. \figcapz}
\newcommand{\figcapmodelone}{{\textsl{\underline{Model 1}}}\; }
\newcommand{\figcapmodeltwo}{{\textsl{\underline{Model 2}}}\; }

\subsection{Directional variation of the cosmographic parameters}
\label{sec:dirvar}
\begin{figure}[ht]
    \centering
    \begin{subfigure}[h]{\textwidth}
    \centering
    \caption{Model 1, Observer 1}
    \includegraphics[width=\linewidth]{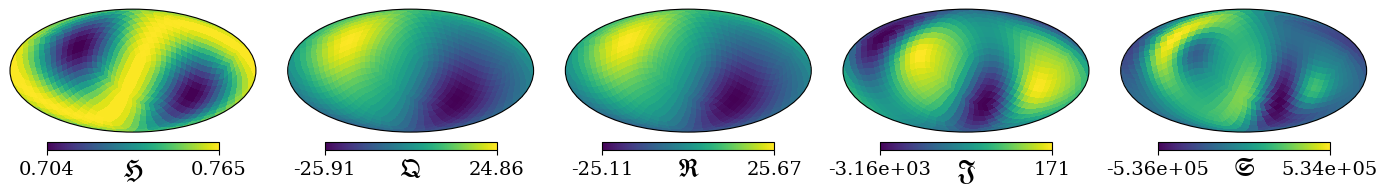}
    \end{subfigure}
    \begin{subfigure}[h]{\textwidth}
    \centering
    \caption{Model 2, Observer 1}
    \includegraphics[width=\linewidth]{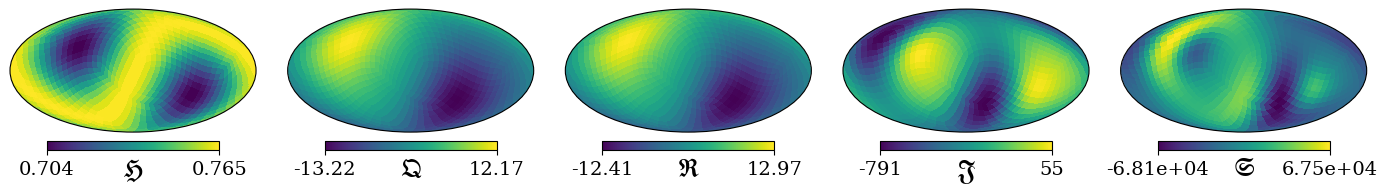}
    \end{subfigure}
    \caption{Variation of the covariant cosmographic parameters (\ref{eq:effh}) and (\ref{eqs:effparams}) for observer 1 in model 1 and model 2.}
    \label{fig:cosmo_plots_all_models_obs1}
\end{figure}

The covariant cosmographic parameters \cref{eqs:effparams} are constructed at the observer location via the kinematical variables of the observer congruence \cref{def:expu}, their covariant derivatives, and the spatial direction of observation on the observer's sky. The sky variation of these parameters is shown in \cref{fig:cosmo_plots_all_models_obs1} for observer~1 (see \cref{tab:obs_params}) in model~1 (see \cref{tab:model_params}).
%
%
In \cref{fig:cosmo_plots_all_models_obs1}, we observe that there is a clear quadrupolar distribution of $\mf{H}$ which is caused by the shear tensor {of the observer congruence}.
We also observe a clear dipolar distribution in the generalised deceleration parameter $\mf{Q}$, agreeing with the stated expectations of \cite{Macpherson:2021gbh}. 
While not visually apparent in the figure, the distribution of $\mf{Q}$ also contains small monopole and octupole contributions, as expected. The dominant dipole in $\mf{Q}$ can be attributed to terms involving $k^\mu\cd{\mu}{\Theta}$, and this dipole is therefore a measure of the relative expansion rate between the (fastly expanding) under-dense and (slowly expanding) over-dense regions of the model. 
The effective curvature parameter has roughly the same structure as $\mf{Q}$. The jerk parameter is dominated by a quadrupole and 16-pole whereas the snap parameter is dominated by a dipole, octupole, and 32-pole. 
Similar variations in the cosmographic parameters are observed for all other observers across both models. 
These observations are all generally expected from the theoretical cosmography expression \cite{Heinesen:2020bej}{, and similar sky patterns were found (up to \ord{3} order in the cosmography) for typical observers in the Einstein Toolkit simulations in \cite{Macpherson:2021gbh}}. 

\subsection{Comparison of ray traced and cosmographic luminosity-distances}
\label{sec:comp}

\begin{figure}[ht]
    \centering
    \begin{subfigure}[h]{\textwidth}
        \centering
        \caption{Model 1 (Obs 1)}
        \includegraphics[width=\linewidth]{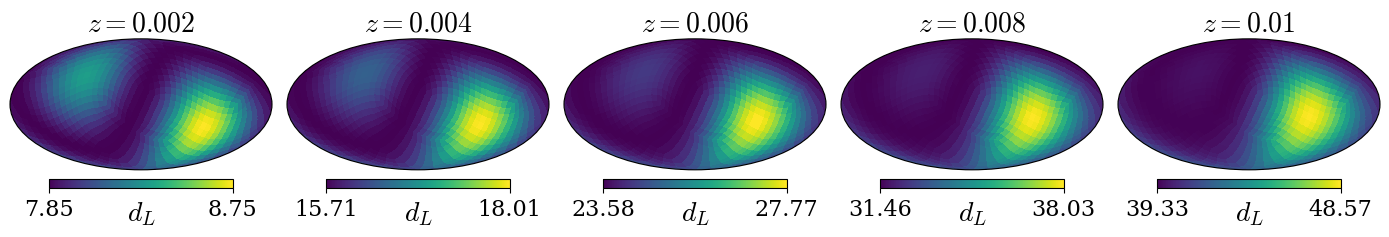}
    \end{subfigure}
    \begin{subfigure}[h]{\textwidth}
        \centering
        \caption{Model 2 (Obs 1)}
        \includegraphics[width=\linewidth]{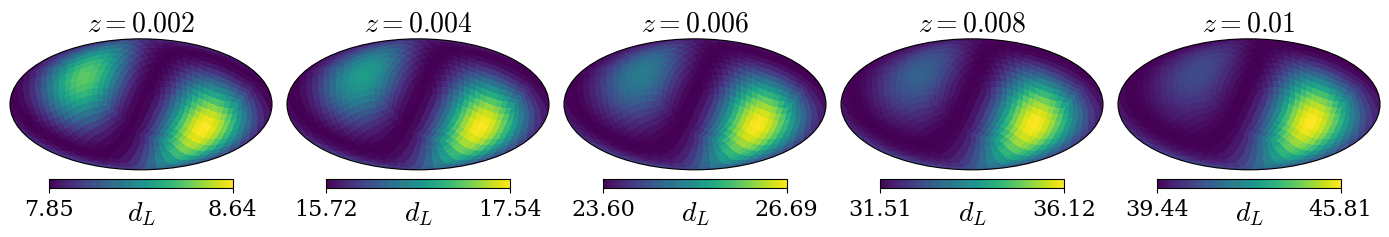}
    \end{subfigure}
    \caption{Variation of the ray traced $d_L$ for observer 1 in models 1 and 2. This is presented as full sky maps for hypothetical sources at five equally spaced redshifts between $z=\dLzmin$ and $z=\dLzmax$, with the observer located at $z=0$.}
    \label{fig:rtdLz_plots_all_models_obs1}
\end{figure}

In \cref{fig:rtdLz_plots_all_models_obs1}, we display the ray traced $d_L$ in models 1 and 2 at five snapshots of redshift. We see that the dominant feature in both plots are a quadrupolar and a dipolar feature, as are the expected main contributions from $\mf{H}$ and $\mf{Q}$ respectively, as discussed in \cref{sec:dirvar}.

\begin{figure}[ht]
    \centering
    \begin{subfigure}[h]{\textwidth}
        \centering
        \caption{Model 1 (Obs 1)}
        \includegraphics[width=\linewidth]{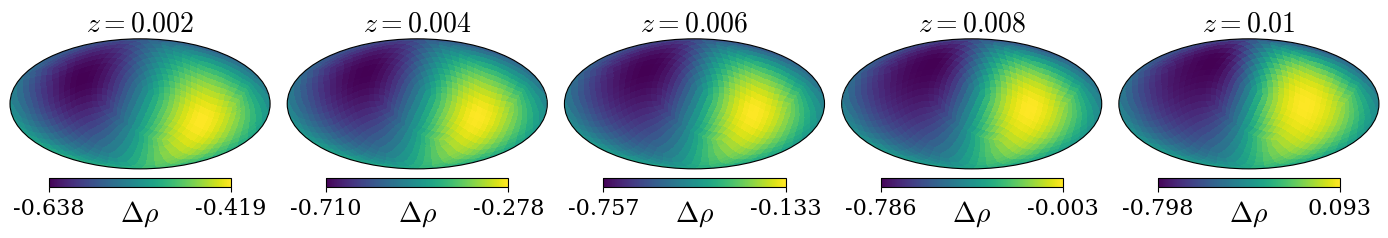}
        \label{subfig:deltarho_obs1_model1}
    \end{subfigure}
    \begin{subfigure}[h]{\textwidth}
        \centering
        \caption{Model 2 (Obs 1)}
        \includegraphics[width=\linewidth]{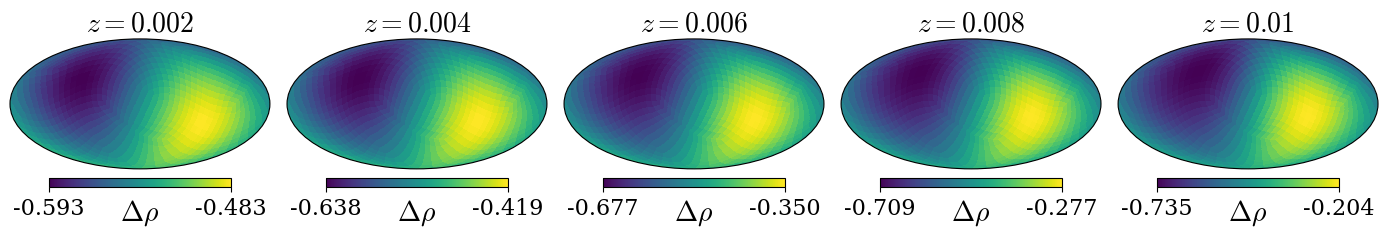}
        \label{subfig:deltarho_obs1_model2}
    \end{subfigure}
    \caption{The density contrast $\Delta\rho$ \cref{eq:deltarho} measured at the redshift position of each hypothetical source for observer 1 in models 1 and 2, for five equally spaced redshifts between $z=0.002$ and $z=0.01$, with the observer located at $z=0$.}
    \label{fig:density_plots_all_models_obs1}
\end{figure}

To assess the estimation ability of the covariant cosmographic expansion at each order, we produce full sky maps of the normalised differences between the cosmographic luminosity distance, $\dLcosmo$, and the ray traced luminosity distance, $\dLRT$, at increasing redshift intervals from a given observer, 
\begin{equation}\label{eq:dLdiff}
    \Delta d_L = \frac{\dLcosmo - \dLRT}{\dLRT}\,.
\end{equation}
\begin{figure}[htb]
    \centering
    \begin{subfigure}[h]{\textwidth}
    \centering
    \caption{\ord{1} order expansion}
    \includegraphics[width=\linewidth]{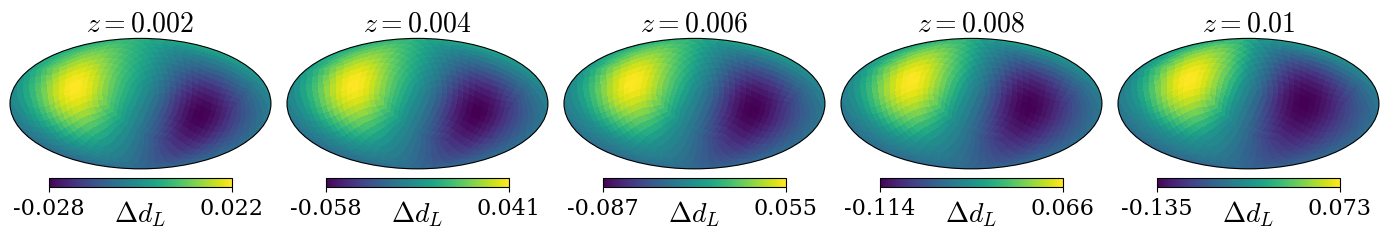}
    \label{subfig:model1_rtdL_diff_obs1_ord1}
\end{subfigure}
\begin{subfigure}[h]{\textwidth}
    \centering
    \caption{\ord{2} order expansion}
    \includegraphics[width=\linewidth]{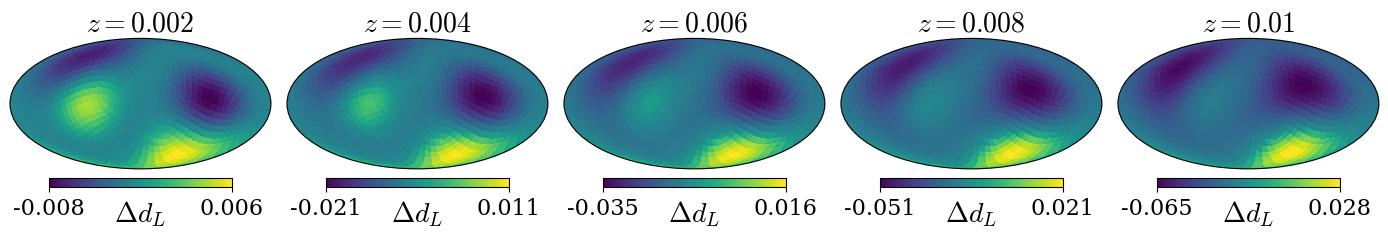}
    \label{subfig:model1_rtdL_diff_obs1_ord2}
\end{subfigure}
\begin{subfigure}[h]{\textwidth}
    \centering
    \caption{\ord{3} order expansion}
    \includegraphics[width=\linewidth]{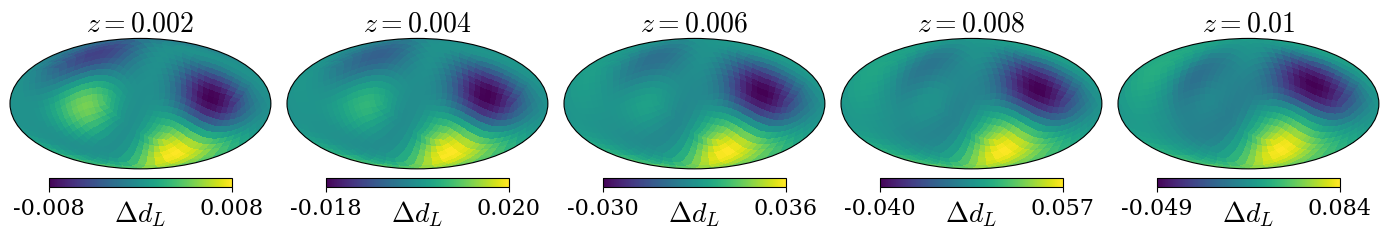}
    \label{subfig:model1_rtdL_diff_obs1_ord3}
\end{subfigure} 
\begin{subfigure}[h]{\textwidth}
    \centering
    \caption{\ord{4} order expansion}
    \includegraphics[width=\linewidth]{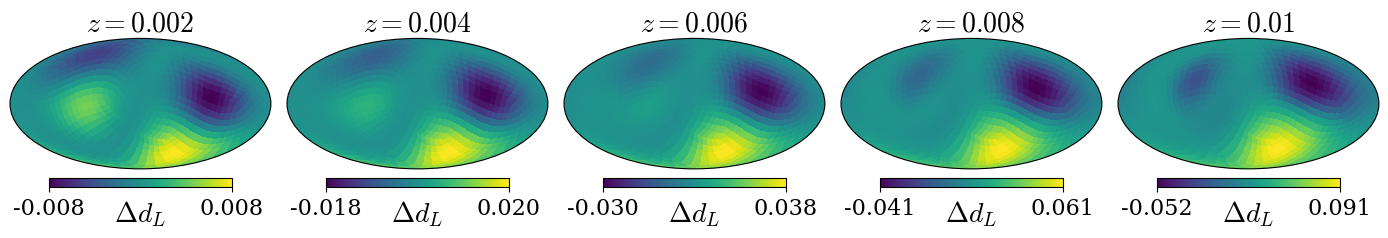}
    \label{subfig:model1_rtdL_diff_obs1_ord4}
\end{subfigure}
    \caption{\figcapmodelone\figcapdLdiffobsone{1}}
    \label{fig:model1_rtdL_diff_obs1}
\end{figure}
\begin{figure}[htb]
    \centering
    \begin{subfigure}[h]{\textwidth}
    \centering
    \caption{\ord{1} order expansion}
    \includegraphics[width=\linewidth]{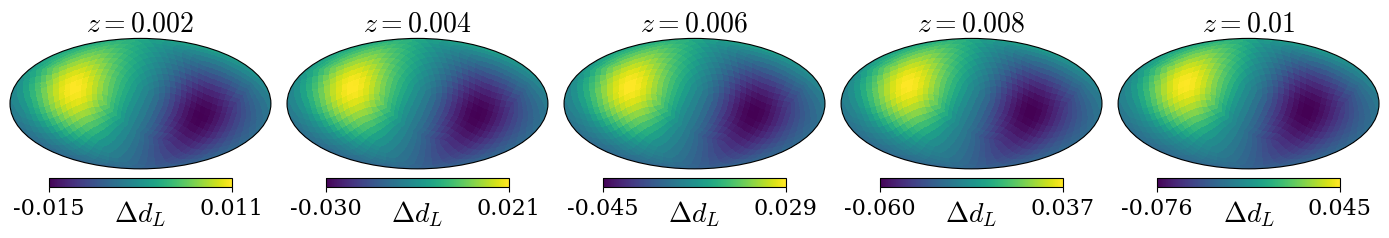}
    \label{subfig:model2_rtdL_diff_obs1_ord1}
\end{subfigure}
\begin{subfigure}[h]{\textwidth}
    \centering
    \caption{\ord{2} order expansion}
    \includegraphics[width=\linewidth]{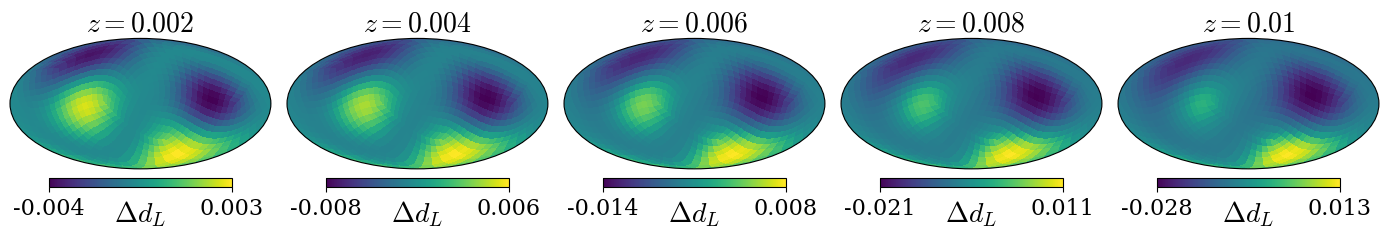}
    \label{subfig:model2_rtdL_diff_obs1_ord2}
\end{subfigure}
\begin{subfigure}[h]{\textwidth}
    \centering
    \caption{\ord{3} order expansion}
    \includegraphics[width=\linewidth]{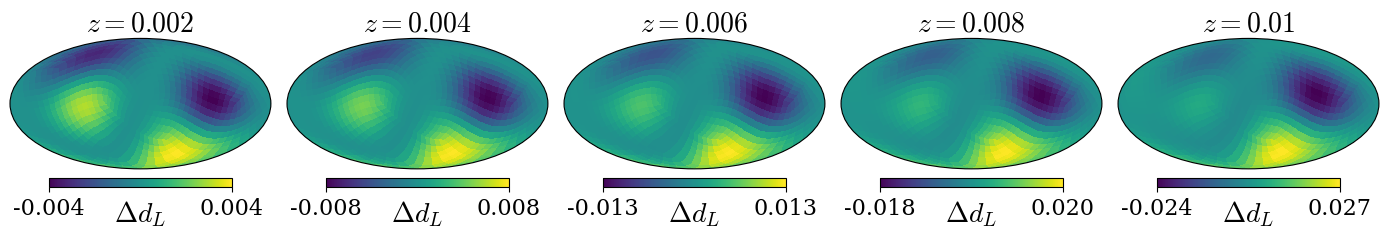}
    \label{subfig:model2_rtdL_diff_obs1_ord3}
\end{subfigure} 
\begin{subfigure}[h]{\textwidth}
    \centering
    \caption{\ord{4} order expansion}
    \includegraphics[width=\linewidth]{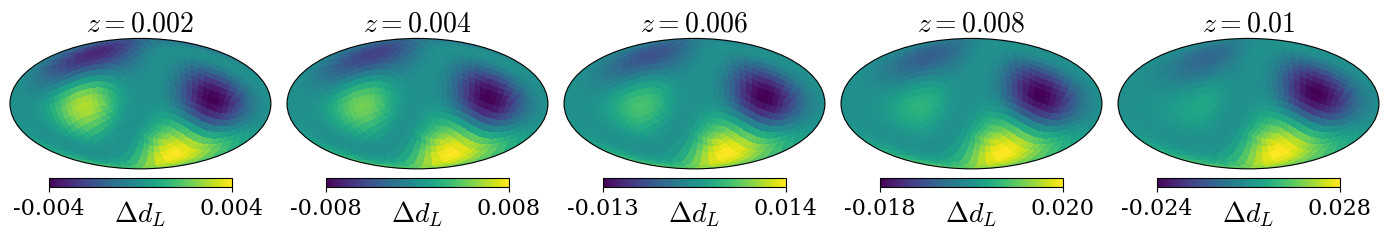}
    \label{subfig:model2_rtdL_diff_obs1_ord4}
\end{subfigure}
    \caption{\figcapmodeltwo\figcapdLdiffobsone{2}}
    \label{fig:model2_rtdL_diff_obs1}
\end{figure}
The {normalized} residuals $\Delta d_L$ for observer 1 for five increasing snapshots of redshift are shown in \cref{fig:model1_rtdL_diff_obs1} for model 1 and \cref{fig:model2_rtdL_diff_obs1} for model 2. 
At \ord{1} order in the cosmographic expansion, 
we see in \cref{subfig:model1_rtdL_diff_obs1_ord1,subfig:model2_rtdL_diff_obs1_ord1} that the leading order term in the cosmography, $z/\mf{H}$, has been successful in accounting for the low-redshift quadrupolar features. Thus, the normalized residuals shown in   \cref{subfig:model1_rtdL_diff_obs1_ord1,subfig:model2_rtdL_diff_obs1_ord1} are dominated by dipolar constributions.
In \cref{subfig:model1_rtdL_diff_obs1_ord2,subfig:model2_rtdL_diff_obs1_ord2} it is in turn seen that these dipoles are accounted for by the \ord{2} order cosmographic term, which is indeed dominated by a dipole through $\mf{Q}$; cf. the discussion in \cref{sec:dirvar}.
We see in \cref{fig:density_plots_all_models_obs1} that the dipoles in density contrast 
\begin{equation}\label{eq:deltarho}
    \Delta\rho = \frac{\rho - \rho_{bg}}{\rho_{bg}}\,,
\end{equation} 
where $\rho_{bg}$ is the background density, align with the dipolar axes seen in \cref{fig:rtdLz_plots_all_models_obs1} and in $\mf{Q}$ in \cref{fig:cosmo_plots_all_models_obs1}. 
This is consistent with the interpretation that the dipole in $\mf{Q}$ is sourced by contrasts in the expansion rate $\Theta$ which is again sourced by density contrasts.
\begin{figure}[ht]
    \centering
    \begin{subfigure}{0.5\textwidth}
        \caption{Observer 1}
        \includegraphics[width=\textwidth]{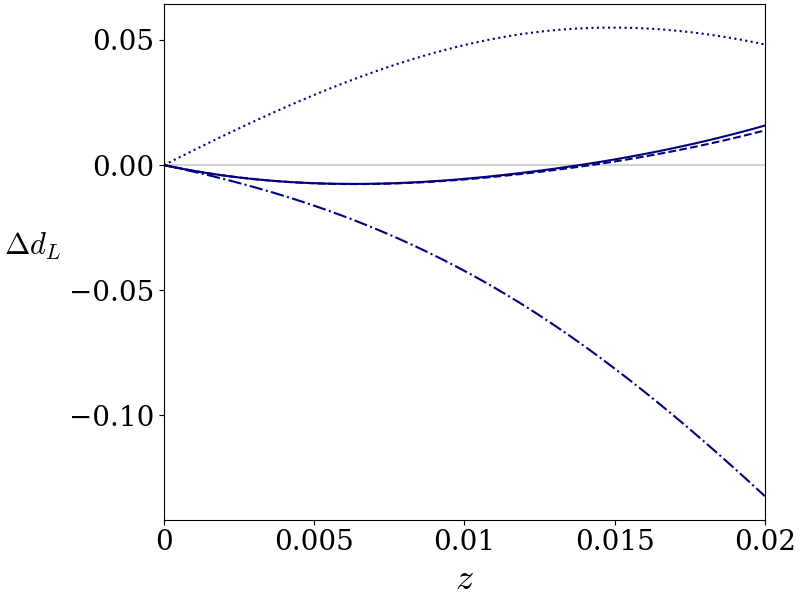} 
    \end{subfigure}%
    \begin{subfigure}{0.5\textwidth}
        \caption{Observer 2}
        \includegraphics[width=\textwidth]{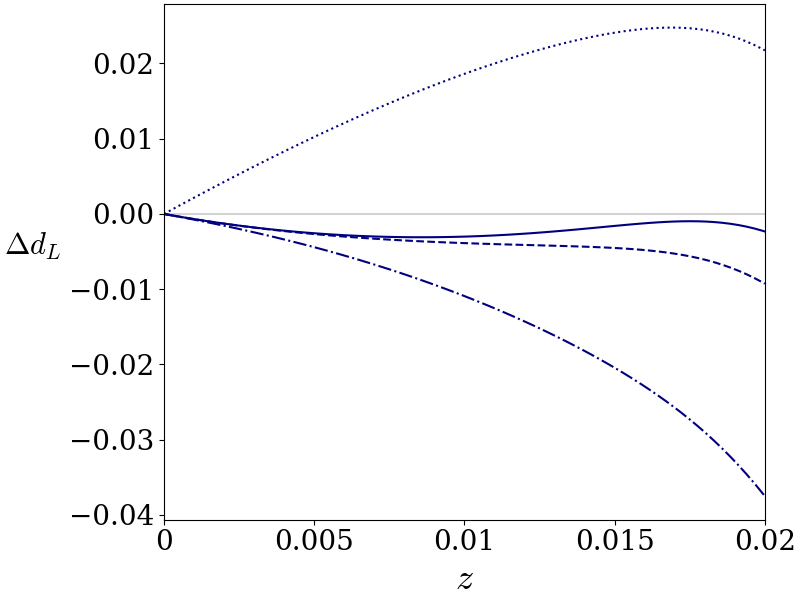}
    \end{subfigure}
    \begin{subfigure}{0.5\textwidth}
        \caption{Observer 3}
        \includegraphics[width=\textwidth]{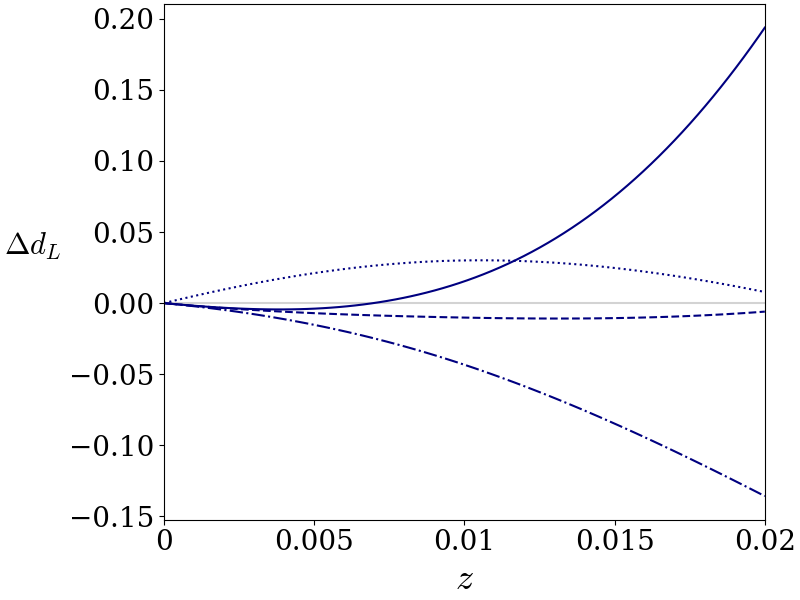}
    \end{subfigure}%
    \begin{subfigure}{0.5\textwidth}
        \caption{Observer 4}
        \includegraphics[width=\textwidth]{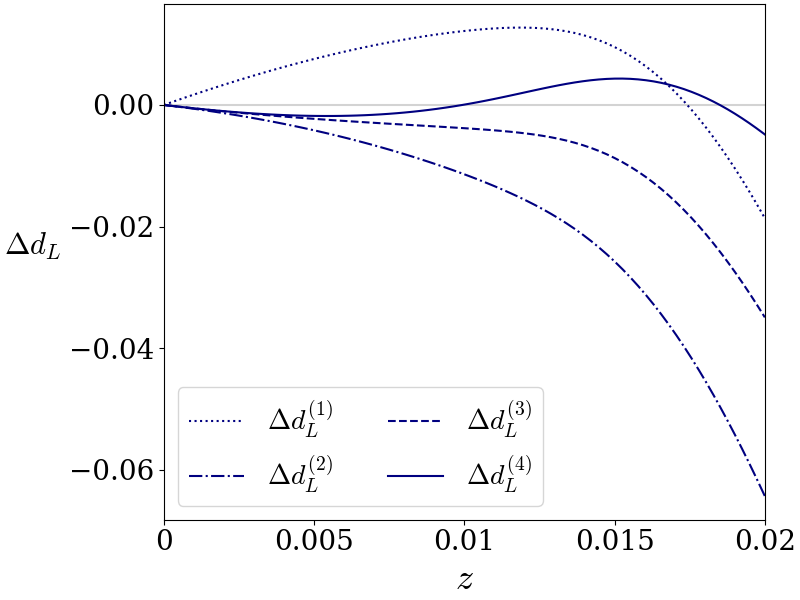}
    \end{subfigure}
    \caption{The convergence of $\DdL$ up to \ord{4} order along a selected test ray with direction coordinates $\qty(\ell, b) = \qty(0.47\pi,0.81\pi)$ on the sky of each observer in model 1.}
    \label{fig:model1_single_rtdL_diff_all_obs}
\end{figure}
\begin{figure}[ht]
    \centering
    \begin{subfigure}{0.5\textwidth}
        \caption{Observer 1}
        \includegraphics[width=\textwidth]{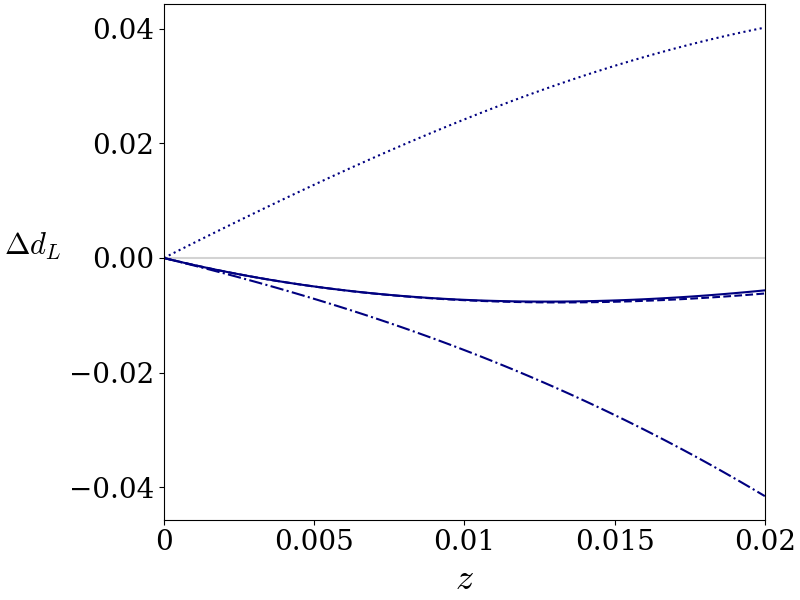} 
    \end{subfigure}%
    \begin{subfigure}{0.5\textwidth}
        \caption{Observer 2}
        \includegraphics[width=\textwidth]{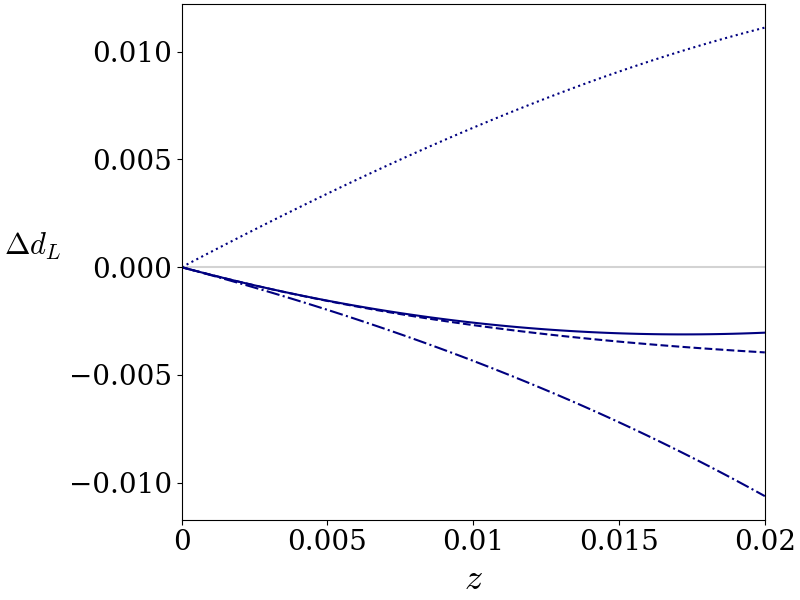}
    \end{subfigure}
    \begin{subfigure}{0.5\textwidth}
        \caption{Observer 3}
        \includegraphics[width=\textwidth]{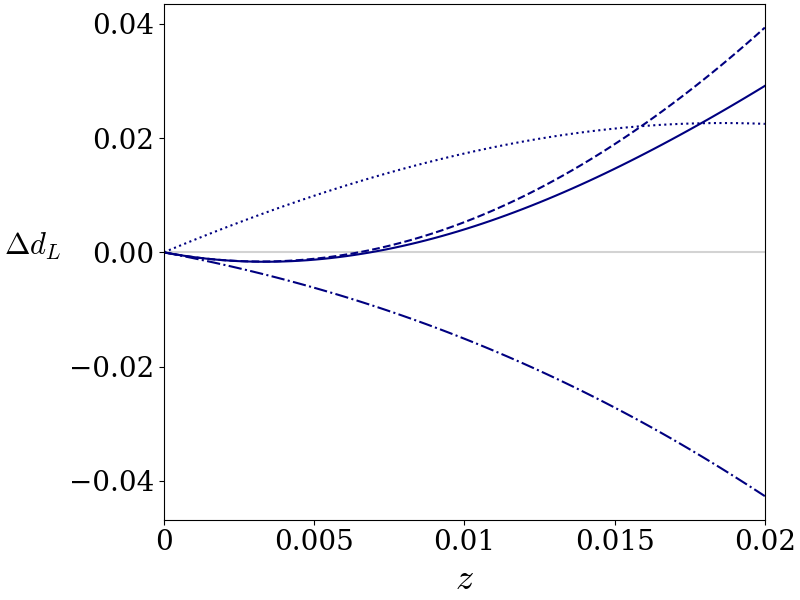}
    \end{subfigure}%
    \begin{subfigure}{0.5\textwidth}
        \caption{Observer 4}
        \includegraphics[width=\textwidth]{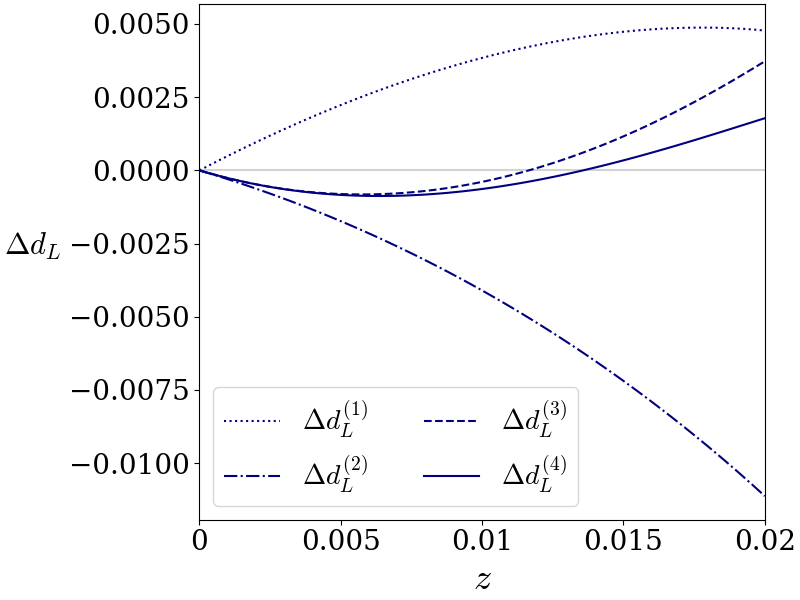}
    \end{subfigure}
    \caption{The convergence of $\DdL$ up to \ord{4} order along a selected test ray with direction coordinates $\qty(\ell, b) = \qty(0.47\pi,0.81\pi)$ on the sky of each observer in model 2.}
    \label{fig:model2_single_rtdL_diff_all_obs}
\end{figure}
In the case of both models, $\Delta d_L$ shows convergence at \ord{2} order for low redshifts without noticeable improvements when including the \ord{3} and \ord{4} order terms. At the larger redshift values shown, up to $z=0.01$, convergence of the cosmographic series by the \ord{4} order is strongly suggested by the similarity of the \ord{3} and \ord{4} order $\Delta d_L$ values, especially for model 2.
The residual $\Delta d_L$ does not vanish, however, even for the lowest redshifts, where $\Delta d_L$ takes values of $\lesssim 0.8\%$ for model 1 and $\lesssim 0.4\%$ for model 2, although the mean of the sky map is very close to 0\% in both models. At $z=0.01$, larger residual values of up to $|\Delta d_L| \lesssim 9 \%$ for model 1 and $\lesssim 3 \%$ for model 2 are observed, again remaining mostly consistent between the \ord{3} and \ord{4} order.
We exclude the numerical resolution in our ray-trace methods as a source of these small observed residuals in appendix~\ref{app:techdetails}. Rather, these likely arise from the small numerical errors associated with solving for the metric function $R(t,r)$ of the Szekeres solution. The cosmography calculations in particular are very sensitive to numerical errors in the solution for the metric components, as the increasing number of derivatives for each cosmographic order tends to amplify errors that may otherwise be modest at the metric level.

Although we have primarily discussed our analysis with regards to observer 1, we have performed the exact same analysis with consistent results for the other 3 observers given in \cref{tab:model_params}. Thus, to provide a snapshot of the full variation of $\DdL$ with redshift across all observer and model combinations, we show $\DdL$ at all orders for a single randomly selected ray in \cref{fig:model1_single_rtdL_diff_all_obs,fig:model2_single_rtdL_diff_all_obs} for model 1 and model 2, respectively. To summarise for observers 2, 3 and 4 at {the lowest redshift} $z\sim0.002$, we find an overall variation in the maximum \ord{4} order residual of $|\DdL|\lesssim0.5\%$ for model 1 and 2. 
{As for observer 1, we {thus} see convergence of the cosmography towards values very similar to the ray traced results, altough the residual is not precicely zero, which we ascribe to numerical errors as for observer 1.}
For these same observers at $z\sim0.01$, we then find an overall variation in the maximum \ord{4} order residual of $|\DdL|\lesssim 2\%$ for both models. 
At redshifts $z\sim0.02$, we then find $|\DdL|\lesssim 20\%$, although most observers/rays perform significantly better with $|\DdL|\lesssim 1\%$ errors. 
{We observe that the \ord{3} and \ord{4} order approximations give roughly consistent results at scales $z\lesssim 0.01$, after which the different orders of the series tend to part.}
These results are consistent with those of observer 1, and {observers 2, 3, and 4 also have} similar full sky distributions to those of observer 1 in \cref{fig:model1_rtdL_diff_obs1,fig:model2_rtdL_diff_obs1}.
We conclude that the \ord{4} order approximation remains very reliable (up to small numerical errors) out to roughly $z\sim 0.01$, after which more terms in the series are needed for a precise (better than $2\%$ or $5\%$ accurate, dependent on the model) approximation. 

\subsection{Assessment of Pad\'e Approximants}

We assessed the convergence behaviour of the $[1/2]$, $[1/3]$, $[2/1]$, $[2/2]$ and $[3/1]$ Pad\'e approximants outlined in \cref{sec:pade}. Additionally, we assessed the $[3/2]$ approximant despite it being technically under-determined for the \ord{4} order expansion of $d _L(z)$ (as $M + N > 4$).  
We find that while the low redshift behaviour of individual rays suggests that some Pad\'e approximants provide a better fit to ray tracing than the standard series expansion, the improvement is largely lost at higher redshifts, with the standard expansion often performing marginally better here. However, a robust conclusion on the performances of Pad\'e approximants beyond third order would require control of the residual numerical errors 
in our analysis discussed in \cref{sec:comp}.

An overall preference for the $[2/1]$ Pad\'e approximant was previously noted by Capozziello \etal \cite{capozziello_high-redshift_2020} for the isotropic FLRW cosmography. The reason that this particular approximant outperformed the $[3/2]$ and $[3/1]$ alternatives is due to degeneracies among the series coefficients when the order $M+N$ of the expansion gets larger.
In particular, increasing the number of terms in the denominator increases the occurrence of poles in the approximation. 
This conclusion agrees with our findings, with the best performance coming from the $[2/1]$ Pad\'e approximant, closely followed by the $[3/1]$ approximant. Moreover, both of these approximants appear free from poles along all discrete lines of sight, which was not the case for all other Pad\'e approximants considered ($[1/2]$, $[1/3]$, $[2/2]$ and $[3/2]$).

Alternative approximations which may be considered in future work include the Chebyshev polynomials investigated in \cite{capozziello_cosmographic_2017}, which were found to generally perform better than the associated higher order Pad\'e and Taylor series expansions. Importantly, the Chebyshev approximant features all higher order terms of the corresponding Taylor series, including those containing the snap parameter $\mf{S}$, within its lower order terms. This is in contrast to range of valid ($M + N \leq 4$) Pad\'e approximants for the \ord{4} order cosmography, which by construction cannot contain the \ord{4} order terms involving the snap parameter. Finally, one might also consider an alternative distance-redshift expansion, such as the inverse series $z(d_L)$ or $z(d_A)$ in conjunction with a suitable alternative approximant, as was done in \cite{adamek_towards_2024,Heinesen:2024npe}.  

\section{Discussion and conclusion}\label{sec:discussion}

In our analysis, we find that the covariant cosmography expansions as applied to our  $\Lambda$-Szekeres models have somewhat low-$z$ radii {within which the \ord{4} order expansion remains a reliable predictor for the exact ray traced luminosity distance--redshift relation, with the \ord{4} order series starting to diverge when approaching the edge of the structure.}  
{This is consistent with previous findigs in numerical relativity simulations and LTB models \cite{Macpherson:2021gbh,Macpherson:2022eve,Sarma:2025yfw}, indicating that low-order cosmographies tend to be reliable only within distances corresponding to the characteristic sizes of the structures in the space-time. }
We note that the level of approximation of the cosmography are highly sensitive to the chosen density profile of the structure. We found that alternative non-analytic parameterisations or choices of parameterisations with more extreme localised spatial gradients than our  canonical Gaussian density profile \eqref{eq:expdelta} generally resulted in less accurate cosmographies as one may have expected. This is again compatible with the findings in \cite{Macpherson:2021gbh,Macpherson:2022eve,Sarma:2025yfw}. 
{The main limitation of our analysis is the pressence of numerical errors in solving for the Szekeres metric function $R(t,r)$ and their propagation to the calculated $d_L$-$z$ relation. Other sources of errors, such as the finite resolution in the ray tracing were shown to be negligible. We expect our overall conclusions to be robust towards improvements in numerical accouracy, and the level of approximaition/converence results that we have presented in this paper for the \ord{4} order cosmography can be taken as conservative estimates.  }

\medskip
Generalised cosmography has been investigated in multiple prior studies, and we comment on those most relevant to our investigations here. 
The investigation presented in this paper may be contrasted to those presented by Modan \& Koksbang in \cite{Modan:2024txm} (Hereafter referred to as MK24), 
who performed a similar analysis using cosmographic formalism for LTB models. 
In particular, they compare the covariant cosmographic expansion to third order in redshift for LTB models to results obtained via general relativistic ray tracing in the models. A key finding was the convergence of the cosmographic series at low radii/redshifts to the ray traced results, which supports our hypothesis that the {small residuals of the cosmography relative to the ray traced results towards very small redshifts} are due to {small} numerical errors {in our solving for the metric rather than signifying inaccouracy in the cosmographic approach.}  

In MK24, the radii of convergence {of the \ord{3} order cosmography} for all models and observers they consider were well within the confines of the LTB structures, which is consistent with the findings of this work. 
Another paper by Sarma \etal on LTB structures \cite{Sarma:2025yfw} arrived at similar conclusions to MK24, with the radii of convergence {of the \ord{3} cosmography} being within the confines of investigated LTB structures. These results, as is also noted in \cite{Sarma:2025yfw}, suggest that care must be taken in interpreting the results when fitting the cosmographic series to data that span multiple structures: We should not expect to arrive at the analytic cosmographic variables derivable from derivatives in a local space-time metric that precisely describe every structure. Rather, the inferred cosmographic parameters should be seen as average parameters that appropriately describe the survey domain. This was also noted in \cite{Macpherson2025theoretical}, where an interpretation of cosmographic paramters (specifically the dipole in the distance-redshift law) were given in terms of \emph{average} space-time variables{, thus extending the convergence of the cosmographic prediction}.

In depth investigations of cosmography as employed to synthetic observers within realistic numerical relativity simulations for cosmology were carried out using the Einstein Toolkit and \emph{gevolution} simulations \cite{Macpherson:2021gbh,Macpherson:2022eve,adamek_towards_2024}. These investigations mainly focused on mapping the large-scale universe above distance scales of $100 - 200 \,\hmpc$. Thus, with our focus on local structures around scales of $100\,\hmpc$ in radii, the investigations in this paper represent complementary investigations to the previously carried out simulation studies.

It is intriquing that a recent analysis of the CosmicFlows4 dataset by Kalbouneh \etal \cite{kalbouneh_anisotropic_2025} utilizing covariant cosmography found significant anisotropies, dominated by a dipole, in our local cosmic neighbourhood $z\lesssim 0.1$ which are in tension with the $\Lambda$CDM model at the $3 \sigma$-level; see also \cite{Boubel:2024cmh} for an alternative approach that arrives at a similar significance level. The results in \cite{kalbouneh_anisotropic_2025} moreover point to axisymmetry of the underlying solution, and thus the axisymmetric Szekeres solutions that we have investigated in this paper may provide a class of models which are interesting for examining data from the local Universe.

\acknowledgments
We thank Timothy Clifton and Sofie Koksbang for helpful discussions, and Pierre Mourier and David Wiltshire for detailed comments on drafts of the manuscript. This work was partly supported by Marsden Fund grant M1271 administered by the Royal Society of New Zealand, Te Ap\=arangi. MH also wishes to thank Zachary Lane, Marco Galoppo and Shreyas Tiruvaskar for insightful discussions and comments. AH also acknowledges the Perren Fund from the University of London and the Astronomy Unit at Queen Mary University of London for their support.

\appendix

\section{Numerical Convergence \& Accuracy}\label{app:techdetails}

Throughout our analysis, we have performed a variety of assessments for the stability and accuracy of both the ray tracing and cosmographic solutions. 
A subtstantial component of our codes is the ray tracing algorithms, which utilise the ODE solvers of the \texttt{DifferentialEquations.jl} package. There are various means for controlling the behaviour of the solver, including the setting of the step size control parameters. One of these parameters is the maximum step size performed by the adaptive `time-stepping' algorithm, which is equivalent to varying the maximum allowed step size of the affine parameter, $\maxstep$, in our numerical routines which integrate the null geodesic equations (which were discussed in \cref{sec:rt_method}). 

To check that the solutions generated by these routines are not variable with this choice of step size, and to gauge if similar control options may affect the results, we have performed our ray tracing analysis scheme for a sample of observer and model combinations while varying $\maxstep$. For each test case, we individually examine 12 rays evenly spread across the observer's sky\footnote{As we use \textsc{HEALPix} to define our inital ray directions, 12 rays correspond to the minimum pixel resolution of \textsc{HEALPix} map.}, by assessing $\DdL$ over the full redshift along a given ray for $0.001\leq\maxstep /\rm{Mpc}\leq8.0$ -- including the $\maxstep/\rm{Mpc}=1.0$ used in our results. Overall, we find that $\Delta d_L$ varies insignificantly across the range of $\maxstep$ tested, indicating that convergence is already achieved when tracing rays with $\maxstep/\rm{Mpc}=1.0$. Examination of the individual rays with different $\maxstep$ yeilds virtually indentical result plots similar to \cref{fig:model1_single_rtdL_diff_all_obs,fig:model2_single_rtdL_diff_all_obs}. Thus, any minute variations are much smaller than the residual $\Delta d_L$ values of order $\order{10^{-2}}$ that we obtained between ray traced and \ord{4} order cosmographic expansion estimates (see \cref{fig:model1_rtdL_diff_obs1} and \cref{fig:model2_rtdL_diff_obs1}). We thus conclude that these residual values should not be due to numerical convergence issues of ray tracing routines themselves. 

In addition to the numerical convergence testing of the ray tracing routines, further performance and stability tests were performed for routines which acquire and take numerical derivatives of certain key functions \eg, those within \cref{eq:evo} and \cref{eq:effh}. This is because the computation of cosmographic parameters in  \cref{sec:distmeasures} contain higher order derivatives of metric functions. 
Overall, we found that our numerical differentiation routines performed well when compared to other candidates explored (\eg, the automatic differentiation procedure we implement for derivatives of $R(t,r)$ with respect to $r$ outperforms manual finite differencing).  

\section{Null vector initialisation procedure}\label{app:nullvecinit}

The local orthonormal basis $\qty{\bs{\basis}_{\hat{\alpha}}}$ can be determined by first considering the line element
\begin{equation}
    \dd{s}^2 = g_{\mu\nu}\dd{x}^\mu\otimes\dd{x}^\nu = \eta_{\hat{\alpha}\hat{\beta}}\,\bs{\omega}^{\hat{\alpha}}\otimes\bs{\omega}^{\hat{\beta}}\,,
\end{equation}
where $\eta_{\hat{\alpha}\hat{\beta}}={\rm diag}\qty(-1,1,1,1)$ and $\qty{\bs{\omega}^{\hat{\alpha}}}$ are the 1-forms dual to the orthonormal basis, such that
\begin{equation}\label{eq:oneform}
    \bs{\omega}^{\hat{\alpha}}\qty(\bs{\basis}_{\hat{\beta}}) = {\omega^{\hat{\alpha}}}_\mu\,{\basis^\mu}_{\!\hat{\beta}} = {\delta^{\hat{\alpha}}}_{\hat{\beta}}\,.
\end{equation} 
By extracting the 1-forms of the Szekeres line element in spherical coordinates (\ref{eq:szek_sph_metric}), we can solve (\ref{eq:oneform}) to obtain the following orthonormal basis 
\begingroup
    \renewcommand*{\arraystretch}{1.5}
    \begin{equation}
        {\basis^\mu}_{\!\hat{\alpha}} = \mqty(
            1 & 0 & 0 & 0 \\
            0 & \zeta & 0 & 0 \\
            0 & \zeta\frac{S'\sin\theta+\mc{N}(1-\cos\theta)}{S} & \frac{1}{R} & 0 \\
            0 & -\zeta\frac{\qty(\partial_\phi\mc{N})\qty(1-\cos\theta)}{S\sin\theta} & 0 & \frac{1}{R\sin\theta})\,,
    \end{equation}
\endgroup
where
\begin{equation}
    \zeta = \sqrt{1-k}\,\qty[R' + \frac{R}{S}\qty(S'\cos\theta + \mc{N}\sin\theta)]^{-1}\,.
\end{equation}
As we have $k^{\hat{\alpha}}_{\rm obs}=\qty(1,-\vu*{n})$ 
in the local orthonormal frame, with the components of $\vu*{n}$ defined in (\ref{eq:localframe}), our initial null vector at the event of observation $k^\mu_{\rm obs}$ is 
\begingroup
    \renewcommand*{\arraystretch}{1.5}
    \begin{equation}
        k^\mu_{\rm obs} = \eval{{\basis^\mu}_{\!\hat{\alpha}}k^{\hat{\alpha}}}_{\rm obs} = \mqty(1 \\
            -\zeta\hat{n}^x \\
            -\zeta\frac{S'\sin\theta+\mc{N}(1-\cos\theta)}{S}\hat{n}^x - \frac{1}{R}\hat{n}^y \\
            \zeta\frac{\qty(\partial_\phi\mc{N})\qty(1-\cos\theta)}{S\sin\theta}\hat{n}^x - \frac{1}{R\sin\theta}\hat{n}^z)_{\rm obs}\,.
    \end{equation}
\endgroup
In the special case of axisymmetry, the above simplifies with $\mc{N} = 0$.

\section{Evolution of the direction vector of the photon congruence} 
\label{sec:derivdirection}

In order to compute higher order derivatives of $\mf{H}$, as needed to obtain the cosmographic expressions, one must necessarily compute the corresponding derivatives of $e^\mu$. The first derivative yields (cf. also \cite{Heinesen:2020bej}),  
\begin{equation}\label{eq:dedl}
     \frac{\rm{D} e^\mu}{ \rm{d} \lambda} = E\qty(e^\mu-u^\mu)\mf{H} - Ee^\nu\qty(\frac{1}{3}\Theta {h^\mu}_\nu + {\sigma^\mu}_\nu)\, , 
 \end{equation} 
where we have used the geodesic equation and the decomposition in \eqref{def:expu}.
The second and third covariant derivatives of the direction vector are respectively given by
    \begin{align*}
        \frac{\rm{D}^2 e^\mu}{ \rm{d} \lambda^2} = &\,E^2\mf{H}\qty(\frac{1}{3}\Theta e^\mu + \qty(e^\nu - u^\nu){{\sigma^\mu}_\nu}) - \frac{1}{3}E\qty(e^\mu\dv{\Theta}{\lambda} + \Theta\frac{\rm{D} e^\mu}{ \rm{d} \lambda}) \\
        & + E\qty(e^\mu - u^\mu)\dv{\mf{H}}{\lambda} - E\qty({{\sigma^\mu}_\nu}\frac{\rm{D} e^\nu}{\rm{d}\lambda} + e^\nu\frac{\rm{D}{\sigma^\mu}_\nu}{\rm{d}\lambda})\,,
    \end{align*} 
    \begin{align*}
        \frac{\rm{D}^3 e^\mu}{ \rm{d} \lambda^3} = &\,E\dv{\mf{H}}{\lambda}\frac{\rm{D} e^\mu}{ \rm{d} \lambda} - 2E\mf{H} \frac{\rm{D}^2 e^\mu}{ \rm{d} \lambda^2}  + \frac{2}{3}E^2\Theta\dv{\mf{H}}{\lambda}e^\mu - \frac{2}{3}E\dv{\Theta}{\lambda}\frac{\rm{D} e^\mu}{ \rm{d} \lambda} - \frac{1}{3}E\Theta \frac{\rm{D}^2 e^\mu}{ \rm{d} \lambda^2}  \\
        & - E{{\sigma^\mu}_\nu} \frac{\rm{D}^2 e^\nu}{ \rm{d} \lambda^2} 
        - E^2\mf{H}\frac{\rm{D} {{\sigma^\mu}_\nu}}{\rm{d} \lambda}u^\nu 
        - 2E\frac{\rm{D} e^\nu}{\rm{d}\lambda}\frac{ \rm{D} {{\sigma^\mu}_\nu}}{\rm{d} \lambda} +\frac{1}{3}E^3\mf{H}\Theta e^\nu {{\sigma^\mu}_\nu} \\
        & + E\qty({e^\mu - u^\mu})\dv[2]{\mf{H}}{\lambda} + E^2\mf{H}\qty({e^\mu - u^\mu})\dv{\mf{H}}{\lambda} + 2E^2 e^\nu {{\sigma^\mu}_\nu}\dv{\mf{H}}{\lambda} \\
        & + E^3\mf{H}e^\nu {{\sigma^\mu}_\gamma{\sigma^\gamma}_\nu}- \frac{1}{3}Ee^\mu \dv[2]{\Theta}{\lambda}- Ee^\nu\frac{\rm{D}^2 {{\sigma^\mu}_\nu}}{\rm{d} \lambda^2}\,,
    \end{align*}
where we have used 
$u^\mu u_\mu = {-}1$, $e^\mu u_\mu = 0$, $k^\mu\nabla_\mu k^\nu =0$ in the derivations, as well as appropriate substitutions of (\ref{eq:kmuE}), (\ref{eq:dEdl}) and (\ref{eq:dedl}).

\section{Metric Derived Quantities} 
\label{sec:metricderived}

Here we list the non-zero Christoffel symbols and the Ricci tensor for the Szekeres model under investigation, along with the derivatives of the dipole function that enter the Ricci tensor.

\subsection{Non-Zero Christoffel Symbols}\label{app:projchris}

\begin{equation*}
    \chris{t}{r}{r} = -\frac{\left(R' \mc{E}-R \mc{E}'\right) \left(\dot{R}' \mc{E}-\dot{R} \mc{E}'\right)}{(k-1) \mc{E}^2}\;, \qquad
    \chris{t}{p}{p} = \chris{t}{q}{q} = \frac{R \dot{R}}{\mc{E}^2}\;, 
\end{equation*}
\begin{equation*}
    \chris{r}{t}{r} = \chris{r}{r}{t} = \frac{\dot{R}' \mc{E}-\dot{R} \mc{E}'}{R' \mc{E}-R \mc{E}'}\;, \qquad \chris{r}{p}{p} = \chris{r}{q}{q} = \frac{(k-1) R}{\mc{E} \left(R' \mc{E}-R \mc{E}'\right)}\;, 
\end{equation*}
    \begin{align*}
        \chris{r}{r}{r} &= \frac{1}{2\mc{E}\left(R' \mc{E}-R \mc{E}'\right)(k-1)}\biggl[\mc{E} \left[\mc{E}' \left(k' R-2 (k-1) R'\right)-2 (k-1) R \mc{E}''\right]\biggr. \\
        &\qquad \biggl.+\mc{E}^2 \left(2 (k-1) R''-k' R'\right)+2 (k-1) R \mc{E}'^2\biggr]\;, \\    
    \end{align*}
\begin{equation*}
    \chris{r}{r}{p} = \chris{r}{p}{r} = \frac{R \left(\qty(\pd_p\mc{E}) \mc{E}'-\mc{E} \qty(\pd_p\mc{E}')\right)}{\mc{E} \left(R' \mc{E}-R \mc{E}'\right)}\;, \qquad
    \chris{r}{r}{q} = \chris{r}{q}{r} = \frac{R \left(\qty(\pd_q\mc{E}) \mc{E}'-\mc{E} \qty(\pd_q\mc{E}')\right)}{\mc{E} \left(R' \mc{E}-R \mc{E}'\right)}\;, 
\end{equation*}
\begin{equation*}
    \chris{p}{t}{p} = \chris{p}{p}{t} = \frac{\dot{R}}{R}\;, \qquad
    \chris{p}{r}{r} = \frac{\left(\mc{E} \qty(\pd_p\mc{E}')-\qty(\pd_p\mc{E}) \mc{E}'\right) \left(R \mc{E}'-R' \mc{E}\right)}{(k-1) R \mc{E}}\;, 
\end{equation*}
\begin{equation*}
    \chris{p}{r}{p} = \chris{p}{p}{r} = \frac{R'}{R}-\frac{\mc{E}'}{\mc{E}}\;, \qquad
    \chris{p}{p}{p} = -\chris{p}{q}{q} = -\frac{\qty(\pd_p\mc{E})}{\mc{E}}\;, \qquad
    \chris{p}{p}{q} = \chris{p}{q}{p} = -\frac{\qty(\pd_q\mc{E})}{\mc{E}}\;, 
\end{equation*}
\begin{equation*}
    \chris{q}{t}{q} = \chris{q}{q}{t} = \frac{\dot{R}}{R}\;, \qquad
    \chris{q}{r}{r} = \frac{\left(\mc{E} \qty(\pd_q\mc{E}')-\qty(\pd_q\mc{E}) \mc{E}'\right) \left(R \mc{E}'-R' \mc{E}\right)}{(k-1) R \mc{E}}\;, 
\end{equation*}
\begin{equation*}
    \chris{q}{r}{q} = \chris{q}{q}{r} = \frac{R'}{R}-\frac{\mc{E}'}{\mc{E}}\;, \qquad
    \chris{q}{p}{q} = \chris{q}{q}{p} = -\frac{\qty(\pd_p\mc{E})}{\mc{E}}\;, 
\end{equation*}
\begin{equation*}
    \chris{q}{q}{q} = -\chris{q}{p}{p} = -\frac{\qty(\pd_q\mc{E})}{\mc{E}}\;.
\end{equation*}

\subsection{Non-Zero Components of the Ricci Tensor}

\begin{equation*}
    R_{tt} = \frac{\Ddot{R} \mc{E}'-\Ddot{R}' \mc{E}}{R' \mc{E}-R \mc{E}'}-\frac{2 \Ddot{R}}{R}\;,
\end{equation*}
    \begin{align*}
        R_{rr} &= \frac{-\left(R' \mc{E}-R \mc{E}'\right)}{(k-1) R \mc{E}^2} \Biggl[\mc{E}' \left(2 \left[\qty(\pd_q\mc{E})^2+\qty(\pd_p\mc{E})^2+1\right]-2 \dot{R}^2-R \Ddot{R}-2 k\right)\Biggr.\\
        &\qquad+ \mc{E} \left(-\left(\pd_{qq}\mc{E}+\pd_{pp}\mc{E}\right) \mc{E}'-2 \left[\qty(\pd_q\mc{E}) \qty(\pd_q\mc{E}')+\qty(\pd_p\mc{E}) \qty(\pd_p\mc{E}')\right]+2 \dot{R} \dot{R}'+R \Ddot{R}'+k'\right)\\
        &\qquad\Biggl.+\left(\pd_{qq}\mc{E}'+\pd_{pp}\mc{E}'\right) \mc{E}^2\Biggr]\;,
    \end{align*}
    \begin{align*}
        R_{pp} &= \frac{1}{2 \mc{E}^2}\Biggl[\frac{R}{R' \mc{E}-R \mc{E}'}\left[\mc{E} \left(-2 \left[\qty(\pd_{pp}\mc{E}) \mc{E}'+\qty(\pd_q\mc{E}) \qty(\pd_q\mc{E}')+\qty(\pd_p\mc{E}) \qty(\pd_p\mc{E}')\right]+2 \dot{R} \dot{R}'+k'\right)\right.\Biggr. \\
        &\qquad\left.+2 \mc{E}' \left(\qty(\pd_q\mc{E})^2+\qty(\pd_p\mc{E})^2-\dot{R}^2-k+1\right)+2 \qty(\pd_{pp}\mc{E}') \mc{E}^2\right]\\
        &\qquad\Biggl.+2 \left(-\qty(\pd_q\mc{E})^2-\qty(\pd_p\mc{E})^2+\mc{E} \left(\pd_{qq}\mc{E}+\pd_{pp}\mc{E}\right)+\dot{R}^2+k-1\right)+2 R \Ddot{R}\Biggr]\;, 
    \end{align*}
    \begin{align*}
        R_{qq} &= \frac{1}{2 \mc{E}^2}\Biggl[\frac{R}{R' \mc{E}-R \mc{E}'}\left[\mc{E} \left(-2 \left[\qty(\pd_{qq}\mc{E}) \mc{E}'+\qty(\pd_q\mc{E}) \qty(\pd_q\mc{E}')+\qty(\pd_p\mc{E}) \qty(\pd_p\mc{E}')\right]+2 \dot{R} \dot{R}'+k'\right)\right.\Biggr. \\
        &\qquad\left.+2 \mc{E}' \left(\qty(\pd_q\mc{E})^2+\qty(\pd_p\mc{E})^2-\dot{R}^2-k+1\right)+2 \qty(\pd_{qq}\mc{E}') \mc{E}^2\right]\\
        &\qquad\Biggl.+2 \left(-\qty(\pd_q\mc{E})^2-\qty(\pd_p\mc{E})^2+\mc{E} \left(\pd_{qq}\mc{E}+\pd_{pp}\mc{E}\right)+\dot{R}^2+k-1\right)+2 R \Ddot{R}\Biggr]\;. 
    \end{align*}

\subsection{Derivatives of the Szekeres dipole function} 
\label{sec:DerDipole}
\[
    \mc{E} = \frac{(p-P)^2 + (q-Q)^2 + S^2}{2S}\;, \]
\[    \mc{E}' = -2\frac{(p-P)P'+(q-Q)Q'-SS'}{(p-P)^2 + (q-Q)^2 + S^2}\,\mc{E} - \frac{S'}{S}\,\mc{E}\;, \]
    \begin{align*}
        \mc{E}'' &= S''\qty(1-\frac{\mc{E}}{S}) + 2\mc{E}\qty(\frac{S'}{S})^2 + \frac{2\qty((p-P)P'+(q-Q)Q')S'}{S^2} \\
        &\qquad +\frac{{P'}^2+{Q'}^2-{S'}^2-(p-P)P''-(q-Q)Q''}{S}\;, 
    \end{align*} 
\[    \pd_p\mc{E} = \frac{p-P}{S}\;, \qquad  
   \pd_p\mc{E}' = -\frac{SP'+(p-P)S'}{S^2}\;, \] 
\[    \pd_q\mc{E} = \frac{q-Q}{S}\;, \qquad
    \pd_q\mc{E}' = -\frac{SQ'+(q-Q)S'}{S^2}\;, \]
\[    \pd_{pp}\mc{E} = \pd_{qq}\mc{E} = \frac{1}{S}\;, \qquad
  \pd_{pp}\mc{E}' = \pd_{qq}\mc{E}' = -\frac{S'}{S^2}\;. \]

\bibliographystyle{JHEP}
\bibliography{biblio.bib}

\end{document}